\documentclass[a4paper,11pt]{article}
\pdfoutput=1 

\usepackage{jheppub} 

\usepackage[T1]{fontenc} 

\usepackage{tikz}
\usetikzlibrary{positioning,decorations.pathmorphing,decorations.markings,arrows}

\tikzset{
mystyle/.style={
  circle,
  inner sep=0pt,
  text width=7mm,
  align=center,
  draw=black,
  fill=white
  }
}

\usepackage{comment} 
\usepackage{empheq}
\usepackage{amsmath}
\usepackage{mathtools}

\title{\boldmath The Radiative Phase Space for the Dynamical Celestial Metric}

\author{Adarsh Sudhakar,}
\author{Amit Suthar}

\affiliation{Institute of Mathematical Sciences,\\ IV Cross Road, CIT Campus, \\ Taramani, Chennai, India \ 600 113 \\}  
 
\affiliation{Homi Bhabha National Institute,\\ Training School Complex, Anushakti Nagar, \\ Mumbai, India \ 400 085, India}

\emailAdd{adarshsu@imsc.res.in}
\emailAdd{amitsuthar@imsc.res.in}

\abstract{ Generalized BMS (gBMS) is the Lie group of the asymptotic symmetries at null infinity, and is proposed to be a symmetry of the quantum S-matrix. Despite much progress in understanding the symplectic structure at null infinity consistent with the  gBMS symmetries, the construction of a radiative phase space where all the physical soft modes and their conjugate partners are identified remains elusive. We construct just such a radiative phase space for linearized gravity by a systematic constraint analysis. In addition, we highlight the difficulties that arise in extending this analysis to the non-linear case. In order to analyze the difficulties we face in extending these ideas to the non-linear setting, we consider a toy model in which we gauge the action of the Weyl scaling in the Weyl BMS group. We find that supertranslations are no longer well-defined symmetries on the reduced phase space of the gauged Weyl, as Weyl scalings do not commute with supertranslations. In this restricted case we obtain the symplectic form and derive the reduced phase space.} 

\begin{document}
\maketitle
\flushbottom

\newcommand{\No}{\overset{o}{N}}
\newcommand{\SNo}{\overset{o}{\mathcal{N}}}
\newcommand{\SNi}{\overset{1}{\mathcal{N}}}
\newcommand{\So}{\overset{o}{\sigma}}
\newcommand{\sC}{\mathfrak{C}}
\newcommand{\sP}{\mathcal{P}}
\newcommand{\p}{\partial}
\newcommand{\bz}{\bar{z}}
\newcommand{\hx}{\hat{x}}
\newcommand{\hy}{\hat{y}}
\newcommand{\hz}{\hat{z}}
\newcommand{\scri}{\mathcal{I}}
\newcommand{\sy}{\mathbb{I}}
\newcommand{\Nob}{\mathfrak{N}}
\newcommand*\widefbox[1]{\fbox{\hspace{1.2em}#1\hspace{1.2em}}}
\newcommand{\no}{\overset{o}{\nabla}}
\newcommand{\bd}{\bar{D}}

\section{Introduction}
\label{sec:intro}
Over the last decade or so, there has been a renewed interest in understanding physics at the boundaries of asymptotically flat space-times. The most well-understood components of the boundary are the null infinities $\mathcal{I}^{\pm}$. Their rich structure encompasses the beautiful discovery of Bondi, Metzner, and Sachs \cite{BMS,Sachs}, that the symmetry group that preserves asymptotic flatness at $\mathcal{I}^{\pm}$ is an infinite dimensional group which is termed the BMS group. It is generated by the \emph{super-translations}, and the Lorentz group on the celestial sphere. Supertranslations are the angle-dependent translations in the null coordinates of $\mathcal{I}^{\pm}$, and ordinary translations are a subgroup of it.

Starting from the seminal work of Barnich and Troessart \cite{bt,bt2}, there have been numerous enhancements of the BMS group obtained by relaxing the boundary conditions on the space-time metric at null infinity. Allowing the celestial metric to fluctuate, while keeping the determinant of the celestial metric fixed, we arrive at the well known extended BMS (eBMS) and generalized BMS (gBMS) groups. In all these extensions and generalizations of the $bms$ algebra, super-translations form an abelian ideal. The Lorentz algebra is isomorphic to the algebra of global CKVs of the celestial sphere, $sl(2,\mathbb{C})$. Hence the original $bms$ algebra is a semi direct sum of super-translations and $sl(2,\mathbb{C})$. $ebms$ algebra extends the Lorentz algebra to include all the local CKVs (meromorphic vector fields) on the celestial sphere. The super-rotations ($ebms$ modulo super-translations) form two copies of the Witt algebra. The other generalization, known as gBMS, enhances the Lorentz group to include all the smooth diffeomorphisms on the celestial sphere. In the case of $gbms$ algebra, super-rotations are generated by smooth vector fields on the sphere.  See \cite{strominger lectures, compere lectures} for a comprehensive review of these developments.

Even though the asymptotic symmetries of asymptotically flat spacetimes have been studied since the 1960s, the main reason for their resurgence in the last decade was the realization that the conservation law associated with the super-translation symmetry is equivalent to the well known factorization of scattering amplitudes in the soft limit\cite{s}, Weinberg soft graviton theorem\cite{Weinberg soft}. The eBMS and gBMS extensions were further consolidation of this connection between the asymptotic symmetries and constraints on the S-matrix, as their conservation law was proved to be equivalent to the sub-leading soft graviton theorem discovered by Cachazo and Strominger \cite{sub, gBMS_Ward, pks, alok miguel 2014}. The new extensions also have an interesting consequence in classical gravity, as the corresponding Noether charges defined at null infinity have been shown to be associated with so-called spin-memory effect \cite{pas}. 

Once we accept the paradigm of studying the group of all symmetries that preserve asymptotic flatness, it leads to further enhancements of the BMS group at null infinity. It was shown in \cite{f} that asymptotic flatness remains preserved even after allowing the area element (celestial metric determinant) to fluctuate. This enhancement of BMS group is known as the Weyl BMS (WBMS) group. In this, an arbitrary Weyl scaling of the celestial metric is allowed along with the superrotations generated by all smooth diffeomorphisms of $S^2$. In this article, we mainly focus on the $gbms$ and $wbms$ algebras.

The space of solutions to Einstein equations for a fixed celestial metric is parameterized by the shear tensor $\sigma_{ab}(u,\hat{x})$. As was shown by Ashtekar and Streubel in the 1980s \cite{as}, one can associate a phase space called radiative phase space, parameterized by the shear tensor $(\sigma_{ab})$ to this space of solutions. In order for the phase space to have a faithful representation of the $bms$ algebra and its enhancements, the radiative phase space was enhanced so that it is parameterized by the shear tensor  $\sigma_{ab}(u,\hat{x})$, and the boundary modes $C(\hat{x}), T_{ab}(\hat{x})$ that transforms inhomogeneously under supertranslations and superrotations respectively. The symplectic structure for this setup splits into the hard sector, parametrized by the shear and News tensors ($N_{ab}(u,\hat{x}) = \p_u\sigma_{ab}$) and the soft sector containing the boundary modes and their conjugate partners. The soft sector is termed so, because the quantized conjugate partners to the boundary modes can be identified with the soft modes of the gravitational field. 

The hard and soft sectors are not independent. The conjugate momenta of the soft modes are related to the hard modes by constraints. Prior to imposing these constraints, we call the phase space \emph{kinematical}. The particular case in which the celestial metric is fixed away from all but one point of the celestial sphere, was analyzed in \cite{am}. In that case, one could solve these constraints and obtain a `physical radiative phase space' in which hard and independent soft modes are identified and which generate a Poisson algebra. However, the situation is far more intricate in the presence of a dynamical celestial metric. The conserved charges corresponding to all the gBMS generators have been derived \cite{mp}, such that their canonical action on the shear field matches with the space-time action of gBMS symmetries. Despite this, the radiative phase space that comprises the independent set of fields obtained by relating the hard shear modes and the soft conjugate momenta to the boundary modes has not yet been derived. 

Our goal in this paper is to initiate a study of just such a radiative phase space at future null infinity $\mathcal{I}^+$ in which the sphere metric is smooth and dynamical. This would be a direct extension of the phase space obtained by He, Lysov, Mitra, and Strominger (HLMS) which admits a faithful action of the BMS group \cite{hlms}.

In particular, we consider two scenarios. Firstly, we obtain the physical radiative phase space for the smooth celestial metric, corresponding to gBMS in the linearized setting. Secondly, we consider the case of HLMS phase space augmented by arbitrary Weyl scalings of the celestial metric, and gauge the Weyl scaling. The main results of the paper and its organization are summarized below.  

\paragraph{Summary and organization of the paper:}

In Section \ref{section the story so far}, we begin with a review of  supertranslations, gBMS, and Weyl-BMS groups. In Section \ref{section 3}, we consider the radiative phase space for the $gbms$ algebra. In Section \ref{linearized section}, we start with the symplectic structure proposed by Campiglia-Peraza in \cite{mp}, suitably adapted to linearized gravity. We then obtain the physical radiative phase space by imposing the appropriate second class constraints via the Dirac bracket analysis\cite{d}. The main result of this analysis is the identification of the physical mode conjugate to the subleading soft News. To the best of our knowledge, this is the first model of a radiative phase space that allows for a smooth and dynamical celestial sphere metric. We then outline the difficulties and subtleties in extending our results to full general relativity \footnote{Note that the Ashtekar-Streubel, as well as HLMS phase spaces obtained in linearized gravity, are isomorphic to the ones obtained in the fully non-linear theory. However, in the cases involving dynamical smooth celestial metric, the radiative phase space structure is considerably simplified due to linearization.}. 

In Section \ref{Weyl section 4}, we move on to the second case of interest. We consider the sub-group of Weyl-BMS which is generated by super-translations and Weyl transformations at $\mathcal{I}^{+}$. In particular we start with the expanded radiative phase space comprising of the News tensors, super-translation soft modes as well as the new modes corresponding to the dynamical area element. The symplectic structure is determined by demanding that it is degenerate along the Weyl orbit, rendering Weyl rescaling to be pure gauge. We then show that the constraints can be solved unambiguously, leading to a radiative phase space. The reduced phase space turns out to be different from the HLMS phase space in the following manner: there is no notion of local hard News in the reduced phase space. We also show explicitly that supertranslations are not well defined on our reduced phase space. The reduced phase space is parameterized by the usual soft modes and the $u$-integrals of arbitrary quantities made out of the hard News. We note that they form a closed algebra.

\subsection*{Notations and Conventions}
\begin{itemize}
\item $\nabla_a$ is the two dimensional covariant derivative compatible with the celestial metric $q_{ab}$. $D_a$ is the Diff-$S^2$ covariant derivative as introduced in \cite{mp}. We also denote by $\mathcal{L}_V$ the Lie derivative along vector field $V$.

\item The identity for the space of symmetric trace-free two tensors is constructed out of the Kronecker delta and the flat metric $\eta$ on the sphere. It is denoted by :
\begin{align}
    \sy^{ab}_{cd} = \delta^{(a}_c\delta^{b)}_d - \frac{1}{2}\eta^{ab}\eta^{cd}\ \ ,  \qquad \sy^{ab}_{cd}\,\sy^{cd}_{mn} = \sy^{ab}_{mn} \ \ , \qquad \sy_{ab,cd} = q_{am}q_{bn}\,\sy^{mn}_{cd} ~. \label{STF I}
\end{align}

\item The symbols $d$ and $\delta$ denote the exterior derivative on spacetime and field-space respectively. 
\item We denote by $X_A$ the Hamiltonian vector field associated with arbitrary phase space variable $A$. For a given symplectic form $\Omega$, it is defined through: 
\begin{align}
    \Omega(Y, X_A) &= \delta A[Y] \ ,
\end{align}
where $Y$ is any vector field on phase space.
\item \noindent We use the following conventions for the Poisson Brackets endowed by a symplectic form $\Omega$ :
\begin{align}
     \{f,g\} = \Omega(X_g,X_f) &= \delta f(X_g) = X_g(f) \ .
\end{align}
\end{itemize}

\section{The Story So Far}
\label{section the story so far}
We are interested in analyzing the null boundaries of the asymptotically flat spacetimes. The null boundary $\scri^+$ is parameterized by  $(u, \hat{x}^a)$. Note that $\hx^a$ are the coordinates on the two sphere, transverse to the radial and temporal directions. We choose the following gauge conditions: $g_{rr} = g_{ra} = 0 = \p_r\, \text{det}(g_{ab}/r^2)$, referred to as Bondi gauge. In this particular gauge choice, the metric at $\scri^+$ takes the following form :
\begin{align}
    \mathrm{d}s^2 = -2\mathrm{d}u^2 - \mathrm{d}u \, \mathrm{d}r + \mathrm{d}\hx^a\mathrm{d}\hx^b\left(r^2\,q_{ab} + r\sigma_{ab} + \hdots \right) + \hdots
\end{align}

Now, we review the phase spaces for supertranslations, gBMS and Weyl. 

\subsection{Review of the HLMS Phase Space}

The symplectic potential at  $\scri^{+}$ for the radiative data (shear and News tensor) was famously derived by Ashtekar and Struebel (AS) \cite{as}, and is given by:
\begin{equation}
    \Theta_{AS} = \int_{\scri} \sqrt{q} \  \sigma^{ab} \delta N_{ab} \ ,
\end{equation}
For a well-defined symplectic structure, the News tensor is required to have the following fall-off conditions: 
\begin{align}
    N_{ab}  \equiv  \p_u\sigma_{ab} \xrightarrow{u \to \pm \infty} \mathcal{O}(\rvert u \rvert^{-1-\epsilon}) \qquad (\epsilon \ >  \ 0) \label{eq:fall} \ .
\end{align}
It leads to the following symplectic form and Poisson bracket:
\begin{align}
    \Omega &= \int_{\mathcal{I}}\sqrt{q}\,\delta N_{ab} \wedge \delta \sigma^{ab} ~,\\
    \big[ N_{ab}(u,\hat{x}) , N_{cd}(u',\hat{y}) \big] &= \frac{1}{2}\p_u\delta(u-u')\,\sy_{ab,cd}\, \frac{1}{\sqrt{q}}\delta^2(\hat{x}-\hat{y}) ~, 
\end{align}
where $\sy_{ab,cd}$ was introduced in \eqref{STF I}.

Super-translation symmetries generate a Hamiltonian action on all the smooth functions defined on the Ashtekar-Streubel phase space ($\Gamma_{AS}$). The corresponding charge (or flux) for super-translation contains a term, linear in soft News tensor ($\int  \mathrm{d}u\,N_{ab}$). However, the soft News or its expected conjugate mode does not exist in $\Gamma_{AS}$. He, Lysov, Mitra, Strominger (HLMS) revisited the canonical derivation of the symplectic structure at $\mathcal{I}$, and showed that there is an enhanced phase space, which includes the constant shear as a boundary mode.\footnote{The HLMS phase space can also be denoted as $\Gamma_{\text{HLMS}}$.} The soft News tensor is its conjugate. HLMS analysis involved extending the AS phase space by adding the soft sector to it, and then imposing physical constraints that relate the soft News as the zero mode of the News tensor \cite{strominger lectures, hlms}.

The constant shear mode is separated from the full shear tensor as follows:
\begin{align}
    & \sigma_{ab}(u,\hat{x}) = \  \overset{o}{\sigma}_{ab}(u,\hx) + C_{ab}(\hx) .\label{constant c mode def} 
    \end{align}
\noindent This split is unique if we demand 
    \begin{align}
     \lim_{u\rightarrow\infty}\big [\overset{o}{\sigma}_{ab}(u,\hx)\ + \ \overset{o}{\sigma}_{ab}(-u,\hx) \big] \equiv \ \ \So{ }^{+}_{ab}(\hat{x}) \ + \So{ }^{-}_{ab}(\hat{x}) = 0  ~.
\end{align}

\noindent $C_{ab}$ is the boundary mode that HLMS introduced into the Ashtekar-Streubel phase  space structure. Making this mode explicit within the original Ashtekar-Streubel symplectic form, one can obtain the following symplectic structure:
\begin{align}
     \Omega &= \int_{\mathcal{I}}\sqrt{q}\,\delta N_{ab} \wedge \delta  \overset{o}{\sigma}{}^{ab} + \int_{S^2}\sqrt{q}\,\delta \left(\int \mathrm{d}uN_{ab}\right) \wedge \delta C^{ab} \\
     &\equiv \int_{\mathcal{I}}\sqrt{q}\,\delta N_{ab} \wedge \delta  \overset{o}{\sigma}{}^{ab} + \int_{S^2}\sqrt{q}\,\delta  \SNo_{ab} \wedge\, \delta C^{ab} ~, \label{raw hlms}
\end{align}
\noindent where $\SNo_{ab}$ is defined through:
\begin{equation}
    \SNo_{ab} \equiv \int  \mathrm{d}u \ N_{ab} ~.
\end{equation}

Let us treat the mode conjugate to the constant shear as independent of the hard News, and denote it by $\No_{ab}$, to allow for the soft ($C_{ab}, \No_{ab}$) and the hard ($N_{ab}$) factorization of the phase space. The physical phase space is now a constraint surface inside $\Gamma_{AS} \times \Gamma_s$. The following are the constraints:
\begin{align}
     \No_{ab}(\hat{x}) \ -  \SNo_{ab}(\hat{x}) &= 0  ~,\\
     \So{ }^{+}_{ab}(\hat{x}) \ + \So{ }^{-}_{ab}(\hat{x}) &= 0 \ .
\end{align}

Before solving the second class constraints and computing the physical brackets, there is a subtlety to be considered. Note that the components of Weyl curvature tensor can be identified as electric and magnetic components, in analogy with the gauge theories. For Christodolou Klainermann spacetimes \cite{Christ} and their appropriate generalizations, the magnetic part of the Weyl tensor vanishes at $\mathcal{I}$. This amounts to the following condition:
\begin{align}
    \lim_{u\to \pm \infty} \nabla_{[a}\nabla^c\sigma_{cb]} \equiv \nabla_{[a}\nabla^c\sigma^\pm_{cb]} = 0  \label{CK condition raw}
    \end{align}

\noindent Since $\ \sigma_{cb}^+ + \sigma_{cb}^- = 2C_{cb} \  \text{and} \    \sigma_{cb}^+ - \sigma_{cb}^- = \,\SNo_{cb} \ $, the general solution to \eqref{CK condition raw} is the following:
\begin{align}\label{eq:CK_soln}
    C_{ab} = -2\nabla_a\nabla_bC^{\text{TF}} \qquad ; \qquad \SNo_{ab} = -2\nabla_a\nabla_b\overset{o}{\mathcal{N}}^{\text{TF}} \ .
\end{align}
\noindent This reduces the degrees of freedom of the boundary modes by half.

Let us consider the symplectic form, \eqref{raw hlms}, treat the soft News ($\No_{ab}$) as independent from the hard News ($N_{ab}$), find the HVFs and kinematical brackets and perform the Dirac analysis to obtain the following physical brackets:
\begin{align}
    \big[ N_{ab}(u,\hat{x}) , N_{cd}(u',\hat{y}) \big] &= \frac{1}{2}\p_u\delta(u-u')\,\sy_{ab,cd}\, \frac{1}{\sqrt{q}}\delta^2(\hat{x}-\hat{y}) \\
    \big[ C(\hat{x}), \overset{o}{N} (\hat{y})\big] &= G(\hat{x},\hat{y}) \ , 
\end{align}
\noindent with $G(\hat{x},\hat{y})$ being the Green's function for the operator $4\nabla^a\nabla^b(\nabla_a\nabla_b)^{\text{TF}}$. 

\subsection{gBMS Phase Space}
Generalized BMS is one generalization of the BMS group, which generalizes the Lorentz algebra to include all the smooth diffeomorphisms of the celestial sphere as superrotations. The following is the action of $gbms$ on the constant shear mode and the celestial metric:
\begin{subequations}
   \begin{equation} \delta_VC_{ab}(\hx) = \left(\mathcal{L}_V-\frac{1}{2}\nabla_mV^m\right) C_{ab} ~, \end{equation}
   \begin{equation}
    \delta_Vq_{ab}(\hx) = \left(\mathcal{L}_V-\nabla_mV^m\right) q_{ab} ~.
    \end{equation}\label{gBMS action on C, q}
\end{subequations}
Note that for arbitrary $V^a$, $\text{det }q_{ab}$ is fixed. Since the action of the $gbms$ algebra deforms the celestial metric $q_{ab}$, the associated phase space must include $q_{ab}$ and its conjugate mode $p^{ab}$. 

The existence of superrotations requires relaxing the boundary conditions on the metric components. For example, the shear tensor $(\sigma_{ab})$, upon the action of superrotations picks up a linear in $u$ contribution: $\sigma_{ab}(u,\hat{x}) \to \sigma_{ab}(u,\hat{x}) + uT_{ab}(\hx)$. The coefficient of the linear in $u$ term is well known as the \emph{Geroch tensor} $(T_{ab})$. If we wish to separate the $T_{ab}$ mode from shear, then the phase space is further expanded to include $T_{ab}$ and its conjugate $\Pi^{ab}$. Note that both $p^{ab}$ and $\Pi^{ab}$ are some functionals of the subleading $(\SNi_{ab})$ and the leading $(\SNo_{ab})$ soft News tensors. The leading and subleading soft News tensors are defined as follows: 
\begin{align}
    \SNo_{ab} = \int \mathrm{d}u\,N_{ab} \ ,  \qquad \SNi_{ab} = \int \mathrm{d}u\,uN_{ab} \ ,\qquad  N_{ab}(u,\hx) \xrightarrow{u\to \pm\infty} |u|^{-2-\delta}~.
\end{align}

Since gBMS corresponds to the subleading soft graviton theorem, it requires new subleading soft News modes in phase space, as shown in \cite{alok miguel 2014, gBMS_Ward}. One way to see how subleading soft News comes into the picture is to see how the $\Gamma_{AS}$ changes by the introduction of the Geroch tensor. 
\begin{align}
 \int_{\mathcal{I}}\delta N_{ab} \wedge \delta \sigma^{ab} &\to \int_{\mathcal{I}}\left(\delta N_{ab} + \delta T_{ab}\right) \wedge \left( \delta \So{ }^{ab} + u\,\delta T^{ab} \right) \\
    &= \int_{\mathcal{I}}\delta N_{ab} \wedge \delta \So{ }^{ab} + \int_{S^2}\delta T_{ab} \wedge \left( \int \mathrm{d}u\,\So{ }^{ab} \right) + \int_{S^2} \delta\left(\int \mathrm{d}u\,uN_{ab} \right) \wedge \delta T^{ab} \nonumber\\
    & = \int_{\mathcal{I}}\delta N_{ab} \wedge \delta \So{ }^{ab} - \int_{S^2}\delta T_{ab}\wedge \delta  \SNi{ }^{ab} +\int_{S^2} \delta  \SNi_{ab}\wedge \,\delta T^{ab} \ .\label{splitting sigma into Tab}
\end{align}
Note that, for simplicity, we are not separating out $\sigma_{ab}$ into the constant shear mode $(C_{ab})$, which would lead to $\int \delta\SNo_{ab} \wedge \delta C^{ab}$ term.

With the relaxed fall-off conditions on the spacetime metric, and the metric on the celestial sphere being dynamical, it is not straightforward to obtain a symplectic structure from first principles. After the initial works by Laddha and Campiglia \cite{alok miguel 2014, gBMS_Ward}, Compere et al \cite{compere} improved the understanding of gBMS phase space by writing the renormalized surface charges (with controlled radial divergences). However, the charges in \cite{compere} are non-integrable $(\Omega(\delta,\delta_V)$ is not an exact form.) The integrable part of the charge can be separated out, and the associated Ward identity turns out to be equivalent to the subleading soft graviton theorem. Charges that differ by a boundary term in the soft sector are equivalent to the correct subleading soft theorem \cite{ashokesen1,ashokesen2}, because the quadratic boundary modes have trivial action on the scattering states. The absence of a canonical hard charge is precisely the angular momentum ambiguity in general relativity. 

The charges (integrable part) in \cite{compere} do not close among themselves and have a 2-cocycle extension. This sits as an obstruction if the superrotations are to be symmetries of the quantum S-matrix. Campiglia and Peraza, in \cite{mp}, obtained the charges exploiting the angular momentum ambiguity (adding quadratic boundary modes), such that there was no 2-cocycle. Demanding the charges close, they were able to write down the following symplectic form compatible with these charges:\footnote{In \cite{mp}, $C_{ab}(u,\hx)$ refers to the shear tensor without separating the constant shear mode. Here, we refer to the constant shear mode as $C_{ab}(\hx)$, and the rest of shear tensor as $\So_{ab}$.} 
\begin{align}
    \Omega = \int_{\scri}& \delta N_{ab} \wedge \delta \So{ }^{ab} + \int_{S^2}\left(\delta \SNo_{ab}\wedge \,\delta C^{ab} + \delta p^{ab}\wedge \delta q_{ab} + \delta \Pi^{ab}\wedge \delta T_{ab}\right) \label{sym form mp} ~,\\
    p^{ab} &= \nabla^{(a}\nabla_c\SNi{ }^{b)c}-\frac{R}{2}\SNi{ }^{ab} + \left( \text{Bilinear in }C,\No \right) \label{pab constraint}~,\\
    \Pi^{ab} &= 2\SNi{ }^{ab} + \left( \text{Bilinear in }C,\No \right) ~. \label{piab constraint}
\end{align}
$R$ here refers to the Ricci scalar curvature corresponding to the celestial metric $q_{ab}$. Note that even though we have the symplectic structure, since there is no hard and soft factorization, we do not get a trivial reduced phase space. This symplectic form shall be our starting point to obtain the reduced phase space in Section \ref{section 3}. 

\subsection{Weyl BMS} \label{sec:Weyl}
Weyl BMS is a further generalization of the gBMS group, where a Weyl rescaling of the metric on the celestial sphere is allowed, along with the smooth diffeomorphisms. It was introduced in \cite{f}, where, the authors separate the scaling action due to Diff-$S^2$, and include it into the action of Weyl. A general infinitesimal asymptotic symmetry transformation $\xi$ is parametrized by a pair of functions $\mathcal{T}$, W and a vector field $V$ on $S^2$. The associated Lie bracket can be written as :

\begin{equation}
    [\xi_{(\mathcal{T}_1,V_1,W_1)},\xi_{(\mathcal{T}_2,V_2,W_2)}] = \xi_{(\mathcal{T}_{12},V_{12},W_{12})} ~,
\end{equation}

\noindent where the parameters $\mathcal{T}_{12}$, $W_{12}$ and $V_{12}$ are given by :

\begin{align}
    \mathcal{T}_{12} &= Y_1 [\mathcal{T}_2] - Y_2[\mathcal{T}_1] + W_2\mathcal{T}_1 - W_1\mathcal{T}_2 ~,\\
    W_{12} &= Y_1[W_2] - Y_2[W_1] ~,\\
    V_{12} &= [V_1,V_2]~.
\end{align}

\noindent The right hand side of the last equation being the Lie bracket of vector fields. \\

The infinitesimal transformations  $\xi_{(\mathcal{T},V=0,W=0)}$ form an abelian ideal and are the usual super-translations. Similarly, the Weyl-part of the Weyl BMS are the transformations $\xi_{(\mathcal{T}=0,V=0,W)}$ and gBMS super-rotations are given by $\xi_{(\mathcal{T}=0,V,W =\frac{1}{2}\nabla_aV^a)}$. The form of the latter is dictated by the property that these transformations preserve the determinant of the celestial sphere metric, as can be seen from the variations written below : 
\begin{align}
    \delta_VC_{ab}(\hx) = \mathcal{L}_VC_{ab}(\hx)\ \ , \qquad  \delta_wC_{ab}(\hx) = -wC_{ab}(\hx) \\
    \delta_Vq_{ab}(\hx) = \mathcal{L}_Vq_{ab}(\hx) \ \ , \qquad  \delta_wq_{ab}(\hx) = -2wq_{ab}(\hx)
\end{align}
The following are the finite Weyl transformation of the radiative modes:
\begin{align}
    N_{ab}(u,\hx) &\to N_{ab}(e^{w(\hx)}u,\hx) ~,\\
    \sqrt{q}(\hx) &\to e^{-2w(\hx)}\sqrt{q}(\hx) ~,\\
    C(\hx) &\to e^{-w(\hx)}C(\hx) ~,\\
    \No(\hx) &\to e^{-w(\hx)}\No(\hx) ~.
\end{align}

In \cite{f}, the authors have performed detailed covariant phase space analysis, and obtained the symplectic structure and charges corresponding to the sphere diffeomorphisms and the Weyl rescalings.

\section{Generalized BMS Reduced Phase Space}
\label{section 3}
Our interest lies in understanding the reduced phase space structure for the gBMS, the Diff($S^2$) algebra. We wish to find the physical brackets between the fundamental fields so that they can be promoted to quantum commutation relations. We start from the Campiglia-Peraza symplectic structure \cite{mp} (See \eqref{sym form mp}):
\begin{align}
    \Omega = \int_{\scri}& \delta N_{ab} \wedge \delta \So{ }^{ab} + \int_{S^2}\left(\delta \SNo_{ab}\wedge \,\delta C^{ab} + \delta p^{ab}\wedge \delta q_{ab} + \delta \Pi^{ab}\wedge \delta T_{ab}\right) ~,\label{omega for full gbms}\\
    p^{ab} &= \nabla^{(a}\nabla_c\SNi{ }^{b)c}-\frac{R}{2}\SNi{ }^{ab} + \left( \text{Bilinear in }C,\No \right) \label{full gbms pab constraint} ~,\\
    \Pi^{ab} &= 2\SNi{ }^{ab} + \left( \text{Bilinear in }C,\No \right)  ~.
\end{align}

Note that this particular symplectic form was not derived from covariant phase space methods \cite{Wald,crnk witten,speranza}. Even though we have this symplectic form, and the action of superrotations on all the fundamental fields, we lack the reduced phase space analysis. The `kinematical' phase space is parameterized by the following conjugate pairs:
\begin{equation*}
\{\So_{ab}\}\cup \{\{\Pi^{ab},T_{ab}\},\{p^{ab},q_{ab}\},\{C,\No\}\}~. 
\end{equation*}
There is no hard and soft factorization, and modes $p^{ab}$ and $\Pi^{ab}$ are non-linearly related to the radiative soft modes. Also, $T_{ab}$ and $q_{ab}$ are not independent, but are related by the following constraint:
\begin{align}
    \nabla^bT_{ab} + \frac{1}{2}\nabla_aR=0 ~.\label{f5 constraint}
\end{align}
Thus, to obtain the reduced phase space for gBMS, we need to find the physical brackets between the radiative modes and the Goldstone modes $(C,T_{ab})$, starting from \eqref{sym form mp}, with \eqref{pab constraint}, \eqref{piab constraint} and \eqref{f5 constraint} as constraints, in addition to the HLMS constraints. However, such a constraint analysis turns out to be difficult. We enumerate some of the obstacles in Section \ref{obstacles section} and in Appendix \ref{appendix B}.  But first, we perform this Dirac analysis for the simplified case of linearized gravity.

\subsection{Linearized Gravity}
\label{linearized section}

The goal of this section is to study the symplectic structure corresponding to the $gbms$ algebra in linearized gravity. Even though we have dynamical $q_{ab}$, it differs only infinitesimally from the plane metric $\eta_{ab}$. The radiative data and the soft News modes are also to be treated perturbatively. We wish to find the physical brackets between the modes of the HLMS phase space supplemented by the mode $q_{ab}$ corresponding to the celestial metric. The angular part of the metric at $\scri^+$ takes the following form:
\begin{align}
    \mathrm{d}s^2 &= \hdots + \mathrm{d}x^a\mathrm{d}x^b\left[r^2\left( \eta_{ab}+ h_{ab}(\hx) \right) + r\left(\sigma_{ab}(u,\hx)+C_{ab}(\hx)\right) + \hdots \right] 
\end{align}

\noindent $N_{ab} = \p_u \sigma_{ab}$, the News tensor, is treated infinitesimally and has the following falloff:
\begin{align}
    N_{ab}(u,\hat{x}) \xrightarrow{u\to \pm\infty} (\text{Constant Mode}) + |u|^{-2-\delta} ~.
\end{align}
The constant mode is the Geroch tensor $T_{ab}$. Note that $\sigma_{ab}$ in this section differs from $\So_{ab}$ from earlier sections by the $uT_{ab}$ contribution. 

The leading and sub-leading soft News tensors can be defined as follows:
\begin{align}
    \overset{o}{\mathcal{N}}_{ab}(\hx) &= \int \mathrm{d}u\,\left( N_{ab}(u,\hx) - \lim_{u'\to\infty} N_{ab}(u',\hat{x}) \right) ~,\\
    \SNi_{ab}(\hx) &= \int \mathrm{d}u\,uN_{ab}(u,\hx) ~.
\end{align}
In the second equation, we are relying on the prescription $\int \mathrm{d}u\, u = 0$. Note that the $\sigma_{ab}$, $C, p^{ab}, \Pi^{ab}$ all are infinitesimal. The celestial metric is written as a perturbation $h_{ab}$ around the plane metric $\eta_{ab}$:
\begin{align}
    q_{ab} = \eta_{ab} +  h_{ab}  \ , \qquad \qquad q^{ab} = \eta^{ab} -  h^{ab} ~.
\end{align}
The condition that the $\text{det }q_{ab} = \text{det } \eta_{ab} = 1 $ translates to the following condition for $h_{ab}$:
\begin{align}
   1= \text{det}(\eta + h) &= \exp \text{tr} \log (\eta +  h) = 1 + \eta^{ab}h_{ab} + \hdots \\
    \Rightarrow \qquad  \eta^{ab} h_{ab} &= 0 ~.
\end{align}
Note that if $q^{ab}$ and $\Pi_{cd}$ are conjugates, then their traces are conjugate modes to each other as well. Hence the conjugate mode $p_{ab}$ to the dynamical metric is tracefree.

\subsubsection{Setup: symplectic form and constraints}
\noindent Since the indices are raised/lowered using the non-dynamical metric $\eta_{ab}$ and all the quadratic terms in radiative data and soft modes are omitted, the symplectic form \eqref{omega for full gbms} simplifies greatly. The constant shear mode $C_{ab}$ and leading soft News tensor simplifies as follows:
\begin{align}
    C_{ab} &= -2\p_a\p_bC^{\text{TF}} ~, \qquad \qquad  \SNo_{ab} = -2\p_a\p_b \No{ }^{\text{TF}} ~.
\end{align}
Let us introduce $\sC = 4(\p_a\p_b)^{\text{TF}}\p_a\p_bC$ for the linearized case. Since the differential operator relating $\sC$ and $C$ is invertible \cite{hlms}, we do not lose any information treating $\sC$ as the fundamental field.\footnote{For the non-linear case, the relation between $\sC$ and $C$ is rather complicated and contains Geroch tensor.} With $\sC$, $\No$, $q_{ab}$ and $p^{ab}$ parameterizing the soft sector, the symplectic form \eqref{omega for full gbms} can be expressed as follows:
\begin{equation} \label{eq:lin_sym}
    \Omega = \int_{\mathcal{I}} \p_u\delta \sigma_{ab}\wedge \delta\sigma^{ab} + \int_{S^2}\left( \delta  \No \wedge\, \delta \sC \,+  \delta p^{ab} \wedge \delta q_{ab}\right) ~.
\end{equation}
Note that for the linearized case, the conjugate mode to $T_{ab}$ is $\Pi^{ab}=2\SNi{ }^{ab}$, and we have absorbed the term $\delta \Pi^{ab}\wedge \delta T_{ab}$ into the Ashtekar-Streubel term, as done in \eqref{splitting sigma into Tab}.

Let us enumerate the constraints that define the physical phase space. The relation between the conjugate mode to $q_{ab}$ and the subleading soft News tensor is a constraint, which we denote $\mathcal{F}_1$. Another constraint, relating the Geroch tensor to the dynamical celestial metric now manifests itself as $\mathcal{F}_2$. The rest of the two constraints are the same as those from the HLMS construction. We collect the constraints below:
\begin{align}
    \mathcal{F}_1^{ab} &= p^{ab} - \left(\partial^a\partial_c\SNi{ }^{bc}\right){}^{\text{STF}} ~,\\
    \mathcal{F}_{2a} &=  \lim_{u \rightarrow \infty}\partial^b N_{ab} + \frac{1}{2} \partial_a\partial_b\partial_ch^{bc}  ~,\\
    \mathcal{F}_{3ab} &=  \sigma^{+}_{ab} + \sigma^{-}_{ab} ~,\\
    \mathcal{F}_{4ab} &=  \SNo_{ab} + 2 \partial_a\partial_b\overset{o}{N} ~.
\end{align}

\subsubsection{Kinematical structure}
The following kinematical brackets can be derived \footnote{We have obtained these brackets by calculating the appropriate Hamiltonian Vector Fields (HVFs). See Appendix \ref{appendix B} for details. Strictly speaking, due to fall off properties of the first term of \eqref{eq:lin_sym} the HVF's derived from it are well defined only up to the regularization $\int_{-\infty}^{\infty} du \ u = 0$ and  $\int_{-\infty}^{\infty} du \ \p_u \left(u\sigma_{ab}\right)= 0$, the latter motivated by the falloffs on $N_{ab}(u)-\lim_{u'\rightarrow \infty}N_{ab}(u')$ and the constraint $\sigma^{+}_{ab}+\sigma^{-}_{ab} = 0$} from \eqref{eq:lin_sym} :
\begin{align}
    \big[ \sigma_{ab}(u,\hat{x})\,,\, N^{cd}(u',\hat{y}) \big] &= \frac{1}{2}\sy^{cd}_{ab}\,\delta(u-u')\delta^2(\hx-\hy)  ~,\\
    \big[\sigma_{ab}(u,\hx)\,,\, \SNo{ }^{cd}(\hy)\big] &= \sy^{cd}_{ab}\,  \delta^2(\hx-\hy)~,\\
    \big[ \sigma_{ab}(u,\hx)\,,\, \SNi{ }^{cd}(\hy)\big] &= \frac{1}{2} u\, \sy^{cd}_{ab} \ \delta^2(\hx-\hy) ~,\\
    \big[ q_{cd}(\hx)\,,\,  p^{ab} (\hy)\big] &=  \sy^{ab}_{cd} \ \delta^2(\hx-\hy) ~,\\
    \big[ \mathfrak{C}(\hat{x}) \,,\,   \overset{o}{N}(\hat{y}) \big] &= \delta^2(\hat{x}-\hy) ~,
\end{align}
where $\sy_{cd}^{ab}$, introduced in \eqref{STF I}, ensures that the trace modes are non dynamical. The rest of the brackets do not survive at the linearized order. For example, all the brackets of $p^{ab}$, except for $[q,p]$ have a radiative field on the RHS, and hence are irrelevant in the linearized setting. For the detailed kinematical structure of the full non-linear gBMS, please refer to Appendix \ref{appendix B}.

Thus, using these kinematical brackets, one can find the non-zero brackets between the constraints:
\begin{align}
    \big[ \mathcal{F}_1^{ab}(\hat{x}) \, , \, \mathcal{F}_{2m}(\hat{y}) \big] &= \left(\frac{1}{2}\p_m\p^a\p^b  - \frac{1}{4}\p^2\p^{(a}\delta^{b)}_m -\frac{1}{8}\eta^{ab}\p^2\p_m  \right) \delta^2(x-y)~,\\
    \big[ \mathcal{F}_3^{ab}(\hx) \, , \, \mathcal{F}_{4cd}(\hy) \big] &= 2 \left(\delta^{(a}_c\delta^{b)}_d - \frac{1}{2}\eta^{ab}\eta_{cd}\right)\delta^2(\hx-\hy) ~.
\end{align}
Here we wish to clarify two things: (i) The Hamiltonian vector fields for the mode $\sigma_{ab}$ are not well defined and we require a prescription for the $[\mathcal{F}_{3ab},\mathcal{F}_{3cd}]$ bracket. Following the reference \cite{am}, we set this bracket to $0$.  (ii) To evaluate $[\mathcal{F}_{2a},\mathcal{F}_{3bc}]$ we need
\begin{align}
    &\left[\lim_{u'\to \infty}N_{ab}(u',\hx)\,,\,  \sigma_{cd}^{+}(\hy)+\sigma^{-}_{cd}(\hy)\right] \nonumber\\
    &\hspace{0.8cm} = \lim_{u'\to \infty} \lim_{u\to \infty} \big[N_{ab}(u',\hx)\,,\, \sigma_{cd}(u,\hy) + \sigma_{cd}(-u,\hy) \big] \sim \lim_{u'\to \infty} \lim_{u\to \infty} \delta(u-u') = 0 ~.
\end{align}

\noindent Another justification for the same is as follows: 
\begin{align}
    \lim_{u'\to \infty} \lim_{u\to \infty} u\left[ N_{ab}(u')\,,\, \sigma^{cd}(u) + \sigma^{cd}(-u) \right] &= \lim_{u'\rightarrow \infty}\bigg [N_{ab}(u')\,,  \int \mathrm{d}u \,\p_u(u\sigma^{cd}(u)) \bigg] \nonumber \\
    &= \lim_{u'\rightarrow \infty}\bigg[N_{ab}(u')\,,\, \SNi{ }^{cd} + \int \mathrm{d}u \ \sigma^{cd}(u)\bigg] \nonumber  \\
    =  \lim_{u' \rightarrow \infty} \ \bigg(\frac{1}{2}\delta^{cd}_{ab} \delta^2(\hx-\hy) &- \frac{1}{2}\frac{\delta}{\delta \sigma_{ab}(u',\hx)}\left(\int \mathrm{d}u \ \sigma^{cd}(u,\hy)\right) \bigg) = 0 ~,
\end{align}
\noindent where we explicitly used the Hamiltonian vector field corresponding to $N_{ab}(u',\hx)$. 

\subsubsection{Inverting the Dirac matrix} 
Getting on with the constraint analysis, we can see that the matrix of brackets of constraints, i.e., the Dirac matrix, is block-diagonal. This simplifies the inversion of the Dirac matrix, which we denote by the letter $M$. The `inverse' of the Dirac matrix will be denoted by $W$ and it is required to satisfy the following equation(s) :
\begin{equation}
    \int \mathrm{d}^2\hy \,M_{IJ}(\hx,\hy)W^{JK}(\hy,\hat{z}) = \delta^2(\hat{x}-\hat{z}) \delta^K_I ~,
\end{equation}

\begin{align}
\int \mathrm{d}^2 \hat{z}    \left(\begin{matrix}
      0 & M_{12}{}^{ab}_m & 0 & 0 \\ 
      M_{21}{}^{ab}_m & 0 & 0 & 0 \\
      0 & 0 & 0 & M_{34}{}_{ab}^{cd} \\
      0 & 0 & M_{43}{}_{ab}^{cd} & 0
    \end{matrix} \right)(\hx,\hat{z})
    &\left(\begin{matrix}
      0 & W_{12}{}_{ab}^n & 0 & 0 \\ 
      W_{21}{}_{cd}^m & 0 & 0 & 0 \\
      0 & 0 & 0 & W_{34}{}_{cd}^{kl} \\
      0 & 0 & W_{43}{}_{cd}^{kl} & 0
    \end{matrix} \right)(\hat{z},\hy) \nonumber\\
    &\qquad\quad = \delta^2(\hx-\hy)\left( 
    \begin{matrix}
    \sy^{ab}_{cd} & 0 & 0 & 0\\
    0 & \delta_m^n & 0 & 0 \\
    0 & 0 & \sy_{ab}^{kl} & 0\\
    0 & 0 & 0 & \sy_{ab}^{kl}
\end{matrix}
    \right) ~.
\end{align}

\noindent Some explanation of the notation used above is in order: the capital letters $I,J,K,$ range from 1-4. The sphere indices that each constraint carries are understood. For instance:
\begin{equation}
M_{IJ} \rvert_{I=1,J=2} = M_{12m}^{ab} = [\mathcal{F}_1^{ab}, \mathcal{F}_{2a}] ~.
\end{equation}

\noindent Let us write the equations that define the inverse for $M_{12}{}^{ab}_m$:
\begin{align}
    \int \mathrm{d}^2z\,\big[ \mathcal{F}_1{}^{ab}(\hx)\,,\, \mathcal{F}_{2m}(\hat{z}) \big]\,W_{21}{}^m_{cd}(\hat{z},\hy) &= \delta^{ab}_{cd}\,\delta^2(\hx-\hy) ~,\\
    \int \mathrm{d}^2z\,\big[\mathcal{F}_{2m}(\hat{x})\,,\, \mathcal{F}_1{}^{ab}(\hat{z})\big]\,W_{12}{}_{ab}^n(\hat{z},\hy) &= \delta^n_m\,\delta^2(\hx-\hy) ~.
\end{align}

\noindent We require the Green's function for the following differential operator:
\begin{subequations}
\begin{equation}
   \left(\frac{1}{2}\p_m\p^a\p^b  - \frac{1}{4}\p^2\p^{(a}\delta^{b)}_m -\frac{1}{8}\eta^{ab}\p^2\p_m  \right)_{(x)} W_{21}{}^m_{cd}(\hx,\hy) = \sP \ \sy^{ab}_{cd} \ \delta^2(\hx-\hy)
\end{equation}
\begin{equation}
    \left(\frac{1}{2}\p_m\p^a\p^b  - \frac{1}{4}\p^2\p^{(a}\delta^{b)}_m -\frac{1}{8}\eta^{ab}\p^2\p_m  \right)_{(x)} W_{21}{}^n_{ab}(\hy,\hx) = - \sP \ \delta^n_m\delta^2(\hx-\hy)
\end{equation}
\end{subequations}

\noindent As these differential operators may have non-trivial kernels, we have included formal operators $\sP$ that project functions onto the subspace wherein they are invertible. In terms of these abstract inverses, we can write down all the non-zero entries of the inverse  Dirac matrix: 
\begin{align}   
    W_{21 ab}^m(\hx,\hy) &= - W_{12 ab}^m(\hy,\hx) ~, \\
    W_{34 ab}^{cd}(\hx,\hy) &=  - W_{43 ab}^{cd}(\hx,\hy) = \frac{1}{2}\delta^{cd}_{ab} \delta^2(\hx-\hy)~.
\end{align}
Before moving on, let us check that the inverse of the Dirac matrix is tracefree $\eta^{ab}W_{21}{}_{ab}^m = 0$.
$$
   \left[\int \mathrm{d}^2x \,W_{21}{}^p_{ab}(\hat{w},\hx) \right]\int \mathrm{d}^2z \, M_{12}{}^{ab}_m(\hx,\hat{z})W_{21}{}_{cd}^m(\hat{z},\hy) = \bigg[\int \mathrm{d}^2x \,W_{21}{}^p_{ab}(\hat{w},\hx) \bigg]\sP\, \sy^{ab}_{cd}\,\delta^2(\hx-\hy) ~,
$$   
\begin{align}
    \Rightarrow \int \mathrm{d}^2z\, W_{21}{}_{cd}^m(\hat{z},\hy)\left[ \int \mathrm{d}^2x \,W_{21}{}^p_{ab}(\hat{w},\hx)M_{12}{}^{ab}_m(\hx,\hat{z}) \right] &= W_{21}{}^p_{ab}(\hat{w},\hy) \sP \sy^{ab}_{cd} \\
    \Rightarrow \quad W_{21}{}_{cd}^p(\hat{w},\hy)\sP &= W_{21}{}^p_{ab}(\hat{w},\hy) \sP \sy^{ab}_{cd} \\
    \Rightarrow \qquad \eta^{cd}W_{21}{}_{cd}^p &= 0~.
\end{align}

\subsubsection{Dirac brackets}
\noindent We are now ready to find the physical brackets between the fundamental modes. We identify the dynamical modes of interest that have kinematical brackets with particular constraints. This is illustrated through the following cartoon : 

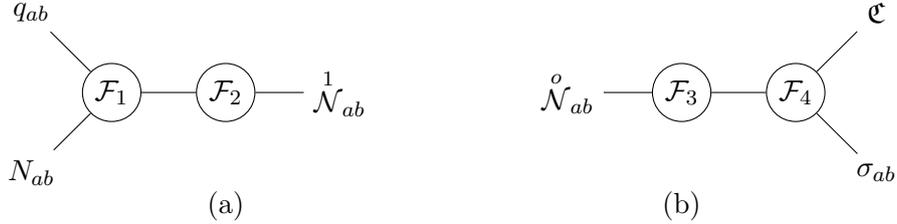
\vspace{0.2cm}

\begin{figure}[ht]
\begin{center}
\begin{tikzpicture}[auto,node distance=1.5cm,main node/.style={circle,draw}]
  \node (1) [mystyle]{$\mathcal{F}_1$};
  \node (2) [right of=1][mystyle] {$\mathcal{F}_2$};
  \draw (1) -- (2);
  \node (3) [above left of=1]{$q_{ab}$};
  \node (4) [below left of=1] {$N_{ab}$};
  \draw (1)--(3);
  \draw (4)--(1);
  \node (5) [right of= 2] {$\SNi_{ab}$};
    \draw (5)--(2);
    \node (10) [right of=2]{};
    \node (11) [right of=10]{};

    \node (7) [right of=11] {$\SNo_{ab}$};
    \node (8) [right of=7][mystyle]{$\mathcal{F}_3$};
    \node (9) [right of=8][mystyle]{$\mathcal{F}_4$};
    \node (12) [above right of=9]{$\sC$};
    \draw (7)--(8);
    \draw (8)--(9);
    \draw (9)--(12);
    \node (13) [below right of=9]{$\sigma_{ab}$};
    \draw (9)--(13);
    \node (14) [below of=2]{(a)};
    \node (15) [below of=8]{(b)};
\end{tikzpicture}
\end{center}
\end{figure}
\noindent An edge between the nodes denoting the constraints indicates that the corresponding entry from the inverse Dirac matrix is non-zero. The kinematical bracket between quantities from the left and right ends of the graph gets corrections from the Dirac analysis. As can be noted from the diagrams, certain brackets are identical to those of the kinematical ones. The only non-vanishing bracket of this kind is:

\begin{equation}
    \big[ N_{ab}(u,\hat{x})\,,\,  N^{cd}(u',\hat{y}) \big]_{*} = \frac{1}{2}\,\sy^{cd}_{ab}\, \p_u\delta(u-u')\delta^2(\hx-\hy).
\end{equation}

\noindent Upon imposing $\mathcal{F}_3$ and $\mathcal{F}_4$, we find that all the brackets of $\SNo_{ab}$ and $\No_{ab}$ turn out to be the same. So, from now on, we don't differentiate between the two. The only non-vanishing bracket involving $\SNo_{ab}$ is:
\begin{equation}
    \big[\SNo_{ab}(\hat{x})\,,\, \sC(\hat{y})\big]_{*} = 2\p_a\p_b^{\text{TF}}\delta^2(\hat{x}-\hat{y}).
\end{equation}

\noindent This matches with the corresponding brackets from HLMS \cite{hlms}. We next consider :
\begin{align} 
    \big[\,h_{ab}(\hx)\,,\, \SNi{ }^{cd}(\hy)\big]_{*} &= - \int \mathrm{d}^2\hat{z} \ \mathrm{d}^2\hat{z}' \ \big[h_{ab}(\hx)\,,\,\mathcal{F}^{ij}_1(\hat{z})\big]\, W^{m}_{12 ij}(\hat{z},\hat{z}')\, \big[\mathcal{F}_{2m}(\hat{z}')\,,\,\SNi{ }^{cd}(\hy)\big]\nonumber \\ &= \frac{1}{2}\sy^{ij}_{ab}\,\sy^{cd}_{mn}\, \frac{\p}{\p y_n} W^{m}_{12 ij}(\hx,\hy) = \frac{1}{2}\sy^{cd}_{mn}\, \frac{\p}{\p y_n} W^{m}_{12 ab}(\hx,\hy). \label{eq:Un_bkt_1}
\end{align}
\noindent Note that in the last equality, we have used $\eta^{ij}W_{12}{}_{ij}^m = 0$.  

\noindent We note from diagram (a):
\begin{align}
    \big[\,q_{ab}(\hx)\,,\, N^{cd}(u,\hy) \big]_* = 0 \qquad \qquad \forall \text{ finite } u \label{qab, Nab bracket}.
\end{align}
However, the two equations, \eqref{qab, Nab bracket} and \eqref{eq:Un_bkt_1} are not in contradiction with each other, because, in general, $[a,\int b] \neq \int [a,b]$. The other non-vanishing bracket is :
\begin{align} 
    \big[&N_{ab}(u,\hx)\,,\, \SNi{ }^{cd}(\hy)\big]_{*}\nonumber\\
    &= \frac{1}{2}\sy^{cd}_{ab}\,\delta^2(\hx-\hy) - \int \mathrm{d}^2\hat{z} \, \mathrm{d}^2\hat{z}' \, \big[N_{ab}(u,\hx)\,,\,\mathcal{F}^{ij}_1(\hat{z})]\, W_{12}{}_{ij}^m(\hat{z},\hat{z}')\, \big[\mathcal{F}_{2m}(\hat{z}')\,,\,\SNi{ }^{cd}(\hy)\big] \\ &= \frac{1}{2}\sy^{cd}_{ab}\,\delta^2(\hx-\hy) - \frac{1}{4} \sy^{ki}_{ab}\,\frac{\p}{\p y_e}\frac{\p^2}{\p x^k \p x_j}W_{12}{}_{ij}^{m}(\hx,\hy)\, \sy^{cd}_{em} ~. \label{eq:Un_bkt_2}
\end{align}
Note that the soft News sector does not decouple completely from the hard News, as was the case in \cite{am}.

\noindent Eliminating the inverse matrix element from \eqref{eq:Un_bkt_1} and \eqref{eq:Un_bkt_2}, we obtain 

\begin{align}
    \bigg[\left(2N_{ab}(u,\hx) + \p^m\p_{(a} h_{b)m}-\frac{1}{2}\eta_{ab}\p^m\p^nh_{mn}\right)\,, \ \SNi{ }^{cd}(\hy)\bigg]_{*} = \sy^{cd}_{ab}\,\delta^2(\hx-\hy)~.\label{conjugate to SNi} 
\end{align}
Thus at the level of linearized gravity, we can find the conjugate mode to $\SNi{ }^{ab}$. The striking thing about this result is that the explicit form of Green's function does not appear in this equation. We only require its existence. Note that since \eqref{conjugate to SNi} holds for all values of $u$, only the constant mode in News, which is exactly the Geroch tensor, is contributing to the Dirac bracket. Had $q_{ab}$ not been dynamical, we would have gotten $2T_{ab}$ to be the conjugate to the subleading soft News tensor, as expected.

\begin{subequations}
\begin{empheq}[box=\widefbox]{align}
 \big[ N_{ab}(u,\hat{x})\,,\, N^{cd}(u',\hat{y}) \big]_{*} &= \frac{1}{2}\,\sy^{cd}_{ab}\, \p_u\delta(u-u')\delta^2(\hx-\hy) \\
\bigg[\big(2N_{ab}(u,\hx) + \sy_{ab}^{mn}\,\p^p\p_{m} q_{np}\big)\,, \ \SNi{ }^{cd}(\hy)\bigg]_{*} &= \sy^{cd}_{ab}\,\delta^2(\hx-\hy) \\
 \big[\sC(\hx) \,,\, \No(\hy) \big]_* &= \delta^2(\hx-\hy)
\end{empheq}
\end{subequations}

\paragraph{Jacobi identities} At the linearized level, all the Dirac brackets are differential operators (or constants) acting on the Dirac delta function, hence the Dirac brackets satisfy Jacobi identities. 

\subsubsection{Celestial plane vs celestial sphere}
In the analysis so far, we have expanded the metric $q_{ab}$ around the flat $\eta_{ab}$ perturbatively. However, we could have done the same analysis with $q_{ab}$ being perturbed around round sphere metric $\overset{o}{q}_{ab}$, for which the Ricci scalar $\overset{o}{R} \,= 2$. The significant difference lies in the kinematical mode conjugate to $h_{ab}$ \eqref{full gbms pab constraint}. The updated constraints would be as follows:
\begin{align}
    \mathcal{F}_1^{ab} &= p^{ab} + \sy^{ab}_{mn}\left( \SNi{ }^{mn} - \no{ }^m \no_c\SNi{ }^{nc} \right) =  p^{ab} + \left( \SNi{ }^{ab} - \no{ }^a \no_c\SNi{ }^{bc} \right)^{\text{STF}} ~, \\
    \mathcal{F}_{2a} &= \lim_{u\to \infty} \no{ }^bN_{ab} + \frac{1}{2} \no_{a}\no{ }^b\no{ }^ch_{bc} ~, \\
    \mathcal{F}_{3ab} &=  \sigma^{+}_{ab} + \sigma^{-}_{ab} ~, \\
    \mathcal{F}_{4ab} &=  \,\SNo_{ab} + 2 \no_a\no_b\overset{o}{N} ~.
\end{align}

\noindent Note that the covariant derivative $\no_a$ is compatible with the round sphere metric $\overset{o}{q}_{ab}$. The non-trivial element of the Dirac matrix is:
\begin{align}
    \big[ \mathcal{F}_1^{ab}&(\hx) \,,\, \mathcal{F}_{2m}(\hy) \big] \nonumber\\
    &= \bigg[\, \frac{1}{2} \no_m\no{ }^a \no{ }^b - \frac{1}{4}\delta^{(a}_m\no{ }^{b)} \no{ }^2 - \frac{1}{8}\overset{o}{q}{ }^{ab}\no_m\,\no{ }^2 + \frac{1}{2}\big(\no{ }^{a}\delta^{b}_m \big)^{\text{STF}} \bigg] \delta^2(\hx-\hy) ~.
\end{align}

\noindent Proceeding as before, the final Dirac brackets are:

\begin{subequations}
\begin{empheq}[box=\widefbox]{align}
 \big[ N_{ab}(u,\hat{x})\,,\, N^{cd}(u',\hat{y}) \big]_{*} &= \frac{1}{2}\,\sy^{cd}_{ab}\, \p_u\delta(u-u')\delta^2(\hx-\hy) \\
\bigg[\bigg(2N_{ab}(u,\hx) - h_{ab}(\hx) + \sy_{ab}^{ik}\,\no_k\no{ }^j q_{ij}\bigg)\,, \ &\SNi{ }^{cd}(\hy)\bigg]_{*} = \sy^{cd}_{ab}\,\delta^2(\hx-\hy) \\
 \big[\sC(\hx) \,,\, \No(\hy) \big]_* &= \delta^2(\hx-\hy)
\end{empheq}
\end{subequations}

\hspace{1cm}

We end this section with a few remarks.
\begin{itemize}
\item We have separated $\overset{o}{N},\, C$ from the hard modes and the reduced phase space admits Poisson brackets between these soft modes along with the usual AS bracket. Hence $\Gamma_{\text{HLMS}}$ is a subspace of the phase space defined in this section.
\item An important distinction of the present analysis from the earlier attempts \cite{mp, am} to derive the symplectic structure on gBMS (eBMS) phase space is the following: In \cite{am} the authors separated the fields in hard and soft sector. In the gBMS case, the soft sector is parametrized by $q_{ab}, T_{ab}$ which are related by a constraint, namely the defining equation for the Geroch tensor. As we have shown,  at least in the linearized gravity, separating hard shear tensor in $\overset{o}{\sigma}_{ab}$ and  $T_{ab}$ is not necessary.  Parameterizing the sub-leading soft sector by sphere metric and its conjugate is sufficient to obtain the reduced phase space in which all the functionally independent conjugate partners can be identified. It is this phase space which should be quantized and may lead us to a more refined understanding of soft vacuua in quantum theory. However, we leave the quantization of this phase space for future work.
\end{itemize}

\subsection{Obstacles in gBMS Phase Space Analysis}
\label{obstacles section}
In this section, we outline some difficulties we faced while undertaking the constraint analysis for $gbms$ algebra. The following is the symplectic form proposed in \cite{mp}, out of which charges that faithfully represent the $gbms$ algebra can be computed:
\begin{equation} \label{eq:gBMS_symp}
\begin{gathered}
    \Omega =  \int_{\mathcal{I}} \sqrt{q} \ \delta N_{ab} \wedge \delta \overset{o}{\sigma}{}^{ab} + \int_{S^2}  \sqrt{q} \ \left( \delta \overset{o}{N} \wedge \ \delta \sC + \delta \Pi^{ab}\wedge \delta T_{ab} +  \delta p^{ab} \wedge \delta q_{ab} \right) \ ,
\end{gathered}
\end{equation}

\noindent where $\sC = \left(-2\nabla_a\nabla_b + q_{ab}\Delta +T_{ab}\right)\left(-2\nabla^a\nabla^b + T^{ab}\right) C$. Furthermore, $p^{ab}$ and $\Pi^{ab}$, the modes conjugate to $q_{ab}$ and $T_{ab}$ respectively, are functions of subleading and leading soft News modes. We include their definitions in the set of constraints:

\begin{align}
    \mathcal{F}^{ab}_1 &= p^{ab} - \left[  \nabla^{a}\nabla_{c}\SNi{ }^{bc}  - \  \frac{R}{2}\SNi{ }^{ab}  + \left( \text{bilinear in} \ C , \ \No \right) \right]  \label{eq:p}~,\\
    \mathcal{F}^{ab}_2 &= \Pi^{ab} -\left[  2\overset{1}{\mathcal{N}}{ }^{ab} + \left( \text{bilinear in} \ C , \ \No \right)\right] \ . \label{eq:Pi_tensor}  
\end{align}
$R$ is the Ricci scalar for the two dimensional celestial sphere. Since we have included the definitions of $p^{ab}$ and $\Pi^{ab}$ as constraints, we can treat them as independent from other soft or hard modes, at the kinematical level. 

Apart from these two, we have two constraints, (almost) the same as the HLMS case. We refer to them as $\mathcal{F}_{3,4}$. The relation between $T_{ab}$ and $q_{ab}$ is referred to as $\mathcal{F}_5$. The same constraints appear in \cite{am}.
\begin{align}
    \mathcal{F}_{3\,ab} &= \,\,\,\SNo_{ab}  + 2 \,[D_aD_b \No]^{\text{TF}} ~,\\
    \mathcal{F}_{4\,ab} &= \,\,\,\So{ }^{+}_{ab} + \So{ }^{-}_{ab} ~,\\
    \mathcal{F}_{5\,a} &= \nabla^b T_{ab} + \frac{1}{2} \nabla_a R \ . \label{f5}
\end{align}

\noindent Here the symbol $D$ denotes the $\text{Diff}(S^2)$ covariant derivative. We have to find the kinematical brackets for this setup and perform the second class constraint analysis to obtain the reduced phase space. However, this problem proved rather difficult. The interested reader may consult Appendix \ref{appendix B} for further details.

\section{Reduced Phase Space for Gauged Weyl BMS}
\label{Weyl section 4}
In this section, we analyze another example of the radiative phase space with relaxed boundary conditions at null infinity. The boundary conditions are such that the celestial metric is fixed up to a conformal factor.  These boundary conditions are `complementary' to those which lead to gBMS symmetries, for which the area form on the celestial sphere is fixed.

It was shown in \cite{f} that such a boundary condition preserves asymptotic flatness. The resulting symmetry group which we denote as ${\cal W}$ is a subgroup of the so-called Weyl-BMS group, discovered in \cite{f}, which is a semi-direct product of the BMS group and the Weyl scaling of the celestial sphere metric. The Weyl-BMS group contains super-translations, celestial diffeomorphisms, as well as Weyl scalings. We focus on the phase space which admits an action of ${\cal W}$, that excludes the area-preserving (celestial) diffeomorphisms. 

Although ${\cal W}$ (as well as Weyl-BMS)  generates an action on the solutions of Einstein's equations, the charge associated with Weyl rescaling does not constrain classical scattering. Moreover, it can be argued that the conservation law for flux $Q_{{\cal W}}$  at ${\cal I}^{+}$, generated by the conformal scaling, is a consequence of the super-translation conservation law.\footnote{We thank Daniele Pranzetti and Laurent Freidel for communicating this result to us.} 

Motivated by this result, we consider the scenario where the asymptotic structure at $\scri$ is fixed up to rescaling freedom of the celestial sphere metric but we then gauge the Weyl
symmetry. This model for a radiative phase space is not physical as in generic radiative space-time the Weyl flux at $\scri^{+}$ need not vanish. However, this example helps us in elucidating subtleties in analyzing the phase space at $\scri$ with a  dynamical celestial sphere metric which we did not encounter in the case of linearized gravity in Section  \ref{linearized section}.

As we show below, the $Q_w = 0$ hypersurface is symplectic and is a direct product of the `soft-sector' parametrized by the super-translation Goldstone mode, its symplectic
partner, and an additional pair of fields associated with the shear field at $\scri$. The soft sector turns out to be isomorphic to the soft sector in $\Gamma_{\text{HLMS}}$ as the only soft modes are $(C,\No)$. The symplectic structure on this hypersurface which is induced by the symplectic structure on $\Gamma_{{\cal W}}$ nicely elucidates the difficulties we face in separating hard and soft degrees of freedom once the celestial metric is a dynamical mode.

\subsection{Weyl Invariance of The CK Condition} \label{magnetic charge vanishing section}
There are two polarizations of the graviton and hence two soft theorems. However, there is only one charge associated with each supertranslation parameter $f(\hat{x})$. The apparent discrepancy is resolved by the \emph{Christodoulou-Klainerman (CK) condition} \cite{Christ}, which relates the positive and negative helicity graviton insertions. CK conditions can be interpreted as the vanishing of the magnetic charge at null infinity \cite{am2016}.\footnote{In analogy with the gauge theories, certain components of the Weyl tensor for the 4-dimensional metric can be identified as electric fields and magnetic fields.} In the presence of a dynamical metric, the magnetic part of the supertranslation charge is as follows:
\begin{align}
    Q_{\mathcal{I}}[\xi_f] = \int_{S^2}f\int \mathrm{d}u\left( \nabla_{[a} \nabla^cN_{cb]} - \frac{1}{2}T_{[a}{}^cN_{cb]}\right) \label{magnetic charge}
\end{align}
Note the presence of an extra term containing Geroch tensor compared to the HLMS setup.  

For the charge \eqref{magnetic charge} to vanish for arbitrary $f(z,\bz)$, the integrand should vanish, hence:
\begin{align}
    \nabla_{[a} \nabla^c \No_{cb]} - \frac{1}{2}T_{[a}{}^c \No_{cb]} = D_{[a} D^c \No_{cb]} = 0 
\end{align}
Here, the $D_a$ are the gBMS covariant derivative, as introduced in \cite{mp}. The general solution to the above constraint is $\No_{ab} = -2D_aD_b\No{}^{TF}$ and hence:
\begin{align}
    D_zD_z\overset{o}{N}{}^{\bz\bz} = D_{\bz}D_{\bz}\overset{o}{N}{}^{zz}
\end{align}
Since $\No_{zz}$ and $\No_{\bz\bz}$ create leading soft gravitons of different helicities, hence the above equation implies that the two polarizations of the leading soft graviton are not independent\cite{hlms}\footnote{Superrotations corresponds to the subleading soft insertions of gravitons, and the insertions of the two helicities are independent. There are two charges corresponding to $V^a \quad a\in [1,2]$ and the electric and magnetic parts for each vector field are identical, hence a total of only 2 independent charges\cite{am2016}.}. 

The counting for the leading soft gravitons remains the same even in the presence of super-rotations, as the CK condition is invariant under super-rotations:
\begin{align}
     \delta_V \left( D_{[a} D^c \No_{cb]} \right) = \left(\mathcal{L}_V + \frac{1}{2}\nabla_aV^a\right) \left(  \nabla_{[a} \nabla^c \No_{cb]} - \frac{1}{2}T_{[a}{}^c \No_{cb]} \right) = 0
\end{align}
Hence, even in the presence of superrotations, the counting for leading soft graviton insertion is still the same.

In order to compute the transformation properties of the magnetic charge under Weyl BMS, we make use of the following identity:
\begin{align}
   \delta_w \vartheta_{a_1\hdots}{}^{b_1\hdots} &= k\,w\,\vartheta_{a_1\hdots}{}^{b_1\hdots}  \quad  \Rightarrow \qquad \delta_w\,D_a\vartheta_{a_1\hdots}{}^{b_1\hdots} = k\,w\,D_a\vartheta_{a_1\hdots}{}^{b_1\hdots}
\end{align}
We make use of the crucial insight that the Weyl weights ($k$ in $\delta_w \vartheta = k \, w\, \vartheta$) and the gBMS weights ($k$ in $\delta_V \vartheta = (\mathcal{L}_V+k\alpha)\vartheta$) coincide for all the quantities.\footnote{This follows from the fact that the $k\alpha$ term in gBMS action: $\delta_V\vartheta = \left(\mathcal{L}_V+k\alpha \right)\vartheta$, upon generalization, leads to the Weyl scalings. $\alpha$ becomes the independent Weyl parameter $w$.} This implies that the gBMS covariant derivative introduced by Campiglia-Peraza in \cite{mp} is also Weyl covariant.\footnote{As an interesting exercise, had the Weyl weight and the gBMS weight been different, we could have constructed a Weyl-Diff-covariant derivative ($\bd$) as follows:
\begin{align*}
     \delta_V \vartheta_{a_1\hdots}{}^{b_1\hdots} &= \left(\mathcal{L}_V+k\alpha\right) \vartheta_{a_1\hdots}{}^{b_1\hdots} \ \ \Rightarrow \qquad \delta_V \bd_a \vartheta_{a_1\hdots}{}^{b_1\hdots} = \left(\mathcal{L}_V+k\alpha\right) \bd_a \vartheta_{a_1\hdots}{}^{b_1\hdots}~,\\
      \delta_w \vartheta_{a_1\hdots}{}^{b_1\hdots} &= l\,w\,\vartheta_{a_1\hdots}{}^{b_1\hdots} \ \ \qquad \quad \ \Rightarrow \qquad \delta_w\,\bd_a\vartheta_{a_1\hdots}{}^{b_1\hdots} = l\,w\,\bd_a\vartheta_{a_1\hdots}{}^{b_1\hdots} 
\end{align*}
This derivative is metric compatible and has the following form:
\begin{align*}
    \bd_a \vartheta_{a_1\hdots}{}^{b_1\hdots} &= D_a \vartheta_{a_1\hdots}{}^{b_1\hdots} + (k-l)A_b\, \vartheta_{a_1\hdots}{}^{b_1\hdots} ~,\\
\text{such that,} \qquad    \delta_wA_b &= \nabla_b w \ \ , \qquad 
    \delta_V A_b = \mathcal{L}_V A_b 
\end{align*} }
As a consequence, the CK condition is invariant under the Weyl scalings:
\begin{align}
    \delta_w \bigg(  D_{[a} D^c \No_{cb]} \bigg) =  0
\end{align}
The usual counting of charges and soft theorems remains unchanged and we can express $C_{ab}$ and $\No_{ab}$ in terms of scalar functions:

\begin{equation} \label{CNO definition}
    C_{ab} \equiv -2D_aD_bC^{\text{TF}}, \quad \No_{ab} \equiv-2D_aD_b\No{}^\text{TF}
\end{equation}

\subsection{Symplectic Structure and Kinematical Brackets}
In this case, a general ansatz for the pre-symplectic form is a sum of the super-translation sector from \cite{am} and a term corresponding to the mode $\sqrt{q}$. Concretely :
\begin{equation}\label{eq:sym_form}
    \Omega = \int_{\scri} \delta N_{ab} \wedge \delta \left(\sqrt{q}\,\So{ }^{ab}\right) + \int_{S^2} \delta  \No \wedge\, \delta \big(\sqrt{q}\,\mathfrak{D}C\big) + \frac{1}{2}\int_{S^2}\delta \Pi \wedge \delta \sqrt{q} ~.
\end{equation}
\noindent The first two terms are the usual terms from $\Gamma_{\text{HLMS}}$, with the last term indicating that the $\sqrt{q}$ is dynamical.\footnote{Note that \eqref{eq:sym_form} reduces to the HLMS symplectic structure once $\sqrt{q}$ is non-dynamical.} Recall that $\mathfrak{D} = 4D_aD_b(D^aD^b)^{TF}$. The presence of the dynamical celestial sphere area element leads to a non-zero Geroch tensor, and hence we have Diff-$S^2$ covariant derivative $D_a$, rather than $\nabla_a$. Also note that since $T_{ab}$ is a functional of $\sqrt{q}$, we are not treating it as an independent boundary mode. The News tensor does not have the Geroch tensor as the constant mode, and hence has the following fall-offs:
\begin{align}
    N_{ab}(u,\hx) \xrightarrow{u\to \pm\infty} |u|^{-2-\delta}~.
\end{align}

An expression for the mode $\Pi$ in terms of the other phase space variables can be obtained from covariant phase space techniques \cite{f}. Here we take a more simplistic approach to fix $\Pi$: we compute it on the hypersurface $\Gamma_\mathcal{W}$ within the phase space, on which we demand that the Weyl flux $Q_\mathcal{W} = \Omega(\delta,\delta_w)$ vanishes identically. The rest of our analysis will be restricted to this hypersurface. The action of Weyl BMS transformations on the modes in the phase space are as follows:
\begin{align}\label{eq:Weyl_var}
    \delta_w N_{ab}(u,\hx) &= w \,u \p_u N_{ab}(u,\hx) \\
    \delta_w \So_{ab}(u,\hx) &= w \left(-\So_{ab} + uN_{ab}\right)(u,\hx) \\
    \delta_w C_{ab}(\hx) &= -w C_{ab}(\hx) \qquad \Longleftrightarrow \qquad \delta_w C = - w C\\
    \delta_w \No_{ab}(\hx) &= -w \No_{ab}(\hx) \qquad \Longleftrightarrow \qquad \delta_w \No = - w \No\\
    \delta_w \sqrt{q}(\hx) &= -2w \sqrt{q}(\hx) \ .
\end{align}
For more details, we refer to \cite{f}.

Demanding that the Weyl flux vanishes puts the following constraint on the phase space variables:
\begin{align} \label{eq:one_find_Pi}
     \Omega(\delta, \delta_w) = \int \mathrm{d}u \ \delta N_{ab}   \sqrt{q}\bigg(\So{}^{ab} &+ u N_{ab} \bigg) - \int \mathrm{d}u \  u\, \p_u N_{ab}\, \delta \left(\sqrt{q}\, \So{}^{ab}\right)  \nonumber\\ +  \No  \ \delta \left(\sqrt{q}\,C^{ab}\right) &+   \delta  \No  \left(\sqrt{q}\, \mathfrak{D} C\right) - \ \delta \Pi   \sqrt{q} - \frac{1}{2w} \delta_w \Pi\, \delta \sqrt{q} = 0 \ .
\end{align}
\noindent Using the falloff conditions on $N_{ab}$, and the relation $\So{}^{+}_{ab} = - \So{}^{-}_{ab}$, we obtain :
\begin{equation} \label{eq:two_find_Pi}
    \delta \left(\int \mathrm{d}u \ u\sqrt{q} N_{ab}N^{ab} +  \sqrt{q}\No \mathfrak{D}C \right) = \sqrt{q} \ \delta \Pi + \frac{1}{2w} \delta_w \Pi \delta \sqrt{q} \ .
\end{equation}
\noindent For the right-hand side to be a total variation, we must have the following:
\begin{equation}\label{eq:cond_Pi}
    \delta_w \Pi = 2w \left(g(z,\bz)f'(\sqrt{q}) + \Pi\right) \ .
\end{equation}
\noindent This implies that $\Pi$ has to be of the form :
\begin{equation} \label{eq:Def_pi}
    \Pi(\hx) = \int \mathrm{d}u \ u N_{ab}N^{ab} +  \No \mathfrak{D}C + g(\hx)f(\sqrt{q}) \ .
\end{equation}

\noindent Substituting it back into the symplectic form \eqref{eq:sym_form}, we see that the term $g(\hx)f(\sqrt{q})$ does not contribute to the symplectic form. We, therefore, omit this term altogether. We have the symplectic form \eqref{eq:sym_form}, from which kinematical Poisson brackets may be derived, and from which the reduced phase space may be obtained by imposing on the kinematic phase space the following constraints:
\begin{align}
    \mathcal{F}_1 &= \ \Pi  -  \No \mathfrak{D}C - \int \mathrm{d}u \ u N_{ab}N^{ab} ~, \label{eq:first_class}\\
    \mathcal{F}_2{}_{ab} &= \ \No_{ab} - \overset{o}{\mathcal{N}}_{ab} ~,\\ 
    \mathcal{F}_3{}_{ab} &= \ \So{}^{+}_{ab} \ + \ \So{}^{-}_{ab} \ .
\end{align}

The symplectic form is degenerate by construction, i.e.,  $ \exists X : \Omega(Y,X) = 0 \ \ \forall\  Y$. 
The $X$ in question is the Hamiltonian vector field for the constraint $\mathcal{F}_1$. This is because $\mathcal{F}_1$ is by construction the Noether charge that generates the Weyl transformations.


At the kinematical level, i.e., prior to the constraints being imposed, we find the brackets using the kinematical HVFs derived from the symplectic form given by \eqref{eq:sym_form}. The second class constraints among these involve the modes $\No$, $\SNo{}^{ab}$, $\So_{ab}$ and $\sqrt{q}$. The non-vanishing brackets amongst these modes are listed below, and these will prove useful when we solve these constraints via the Dirac procedure.
\begin{align} \label{eq:kin_start}
    \big[N^{ab}(u,\hat{x})\,,\,N_{cd}(u',\hat{y})\big] &= \frac{1}{2} \,\p_u \delta(u-u')\, \sy^{ab}_{cd} \ \frac{1}{\sqrt{q}} \delta^2(\hx-\hy) ~,\\
    \big[\SNo{ }^{ab}(\hat{x})\,,\,\So_{cd} (u',\hat{y})\big] &= - \sy^{ab}_{cd} \ \frac{1}{\sqrt{q}}\delta^2(\hat{x}-\hat{y}) ~,\\
\big[ \No(\hat{x})\,,\,C(\hat{y})] &= -\,\frac{1}{\sqrt{q}}G(\hx,\hy) ~,\\ 
    \big[ \So_{ab}(u,\hx) \, ,\, \Pi(\hy) \big] &= \So_{ab}\,\frac{1}{\sqrt{q}}\delta^2(\hx-\hy) ~,\\
    \big[\sqrt{q}(\hx)\,,\,\Pi(\hy)] &= 2 \delta^2(\hx-\hy) ~.
\label{eq:kin_end}
\end{align}

\subsection{Physical Phase Space}
We have a dynamical system with Weyl scalings along $\scri$ as the gauge redundancy. Gauging the Weyl transformations implies that we consider the constraint hypersurface $\mathcal{F}_1 \sim 0$ and quotient it out by the Weyl transformations. It can be checked that this is a consistent restriction as the HVF generating the Weyl action is tangential to the $\mathcal{F}_1 \sim 0$ hypersurface. We present two ways to `solve' such a system, by gauge fixing, and by finding the explicit reduced phase space.

\subsubsection{Gauge fixing} \label{gauge_fix}
One way to `solve' the first class constraint is by introducing another constraint that has a non-zero bracket with the first class constraint, rendering it second class\cite{Gitman,sundermeyer}. To begin with, we have a total of $5\times \infty$ constraints, out of which $1\times \infty$ are first class. $\mathcal{F}_1$ is the first class constraint and is the generator of the Weyl scalings. The new constraint is the gauge fixing constraint. One natural gauge fixing constraint for our current context is $\sqrt{q}=1$\footnote{Other possibilities are $C$ or $\No$ = $\{-1,0,+1\}$. Note that the Weyl can not switch the sign, and if for instance $C=0$ in some region, Weyl action will keep it zero. Hence such a gauge fixing will only be partial.}. We have thus the following second-class constraints to solve:
\begin{align}
    \mathcal{F}_1 &= \ \Pi  -  \No \mathfrak{D}C - \int \mathrm{d}u \ u N_{ab}N^{ab} ~,\\
    \mathcal{F}_2{}_{ab} &= \ -2[D_aD_b \No]^{\text{TF}} - \overset{o}{\mathcal{N}}_{ab}  ~,\\
    \mathcal{F}_3{}_{ab} &= \ \So{}^{+}_{ab} \ + \ \So{}^{-}_{ab} \ , \\
    \mathcal{F}_4 &= \sqrt{q}-1 \ .
\end{align}

We can find the kinematical brackets among the constraints using the kinematical brackets from earlier. Using the HVF:
\begin{align}
    X_{\int \mathrm{d}u\,uN_{ab}N^{ab}} = \frac{1}{\sqrt{q}}\int \mathrm{d}u\,uN_{ab}\frac{\delta}{\delta \So_{ab}} + 2\int \mathrm{d}u\,uN_{ab}N^{ab}\frac{\delta}{\delta \Pi} ~, \label{kleven HVF}
\end{align}
\noindent we can check that:
\begin{align}
    \left[ \int \mathrm{d}u\,uN_{ab}N^{ab} \,,\, \SNo_{mn} \right] &= 0 = \left[ \int \mathrm{d}u\,uN_{ab}N^{ab}\,,\, \mathcal{F}_{3mn}\right] ~.
\end{align}
Hence, the Dirac matrix is as follows\footnote{Note that the antisymmetry of $\big[\mathcal{F}_1\,,\,\mathcal{F}_1\big]$ rules out terms proportional to $\delta^2(x-y)$. One can check that the rest of the terms vanish, and hence $\big[\mathcal{F}_1\,,\,\mathcal{F}_1\big]=0$.} :
\begin{align}
   \left( \begin{matrix}
    0 & M_{12ab}(\hx,\hz) & 0 & -\delta^2(\hx-\hz) \\
    M_{21ab}(\hx,\hz) & 0 & \frac{2}{\sqrt{q}}\sy_{ab,cd}\delta^2(\hx-\hz) & 0 \\
    0 & \frac{2}{\sqrt{q}}\sy_{ab,cd}\delta^2(\hx-\hz) & 0 & 0 \\
    \delta^2(\hx-\hz) & 0 & 0 & 0
    \end{matrix} \right) ~.
\end{align}
The inverse of the constraint matrix is defined as follows: (Note that $I,J,\hdots$ refer to the constraint number $[1,2,3,4]$ along with the sphere indices.)
\begin{align}
 \int \mathrm{d}^2\hat{z} \,M_{IJ}(\hx,\hat{z})W^{JK}(\hat{z},\hy) &= \delta^2(\hat{x}-\hat{y}) \delta^K_I~.
\end{align}
The explicit inverse of the Dirac matrix is as follows:
\begin{equation}
\begin{gathered}
     \int \mathrm{d}^2\hz\left( \begin{matrix}
    0 & M_{12mn} & 0 & -\delta^2_{\hx,\hz} \\
    M_{21ab} & 0 & \frac{2}{\sqrt{q}}\sy_{ab,mn}\delta^2_{\hx,\hz} & 0 \\
    0 & -\frac{2}{\sqrt{q}}\sy_{ab,mn}\delta^2_{\hx,\hz} & 0 & 0 \\
   \delta^2_{\hx,\hz} & 0 & 0 & 0
    \end{matrix} \right) ~.\hspace{5cm} \\
    \hspace{1.5cm} \left( \begin{matrix}
    0 & 0 & 0 & \delta^2_{\hz,\hy} \\
    0 & 0 & -\frac{\sqrt{q}}{2}\sy^{mn,cd}\delta^2_{\hz,\hy} & 0 \\
    0 & \frac{\sqrt{q}}{2}\sy^{mn,cd}\delta^2_{\hz,\hy} & 0 & -\frac{\sqrt{q}}{2}M_{21}{}^{mn} \\
    -\delta^2_{\hz,\hy} & 0 & -\frac{\sqrt{q}}{2}M_{12}{}^{cd} & 0
    \end{matrix} \right) = 
    \left( \begin{matrix}
    1 & 0 & 0 & 0 \\
    0 & \sy^{cd}_{ab} & 0 & 0 \\
    0 & 0 & \sy^{cd}_{ab} & 0 \\
    0 & 0 & 0 & 1 
    \end{matrix} \right)\delta^2_{\hx,\hy} ~.
\end{gathered}
\label{inverse MW}
\end{equation}

Let us look at the corrected brackets. 
\begin{align}
    \big[\varphi,\alpha\big]_* &= \big[\varphi,\alpha\big] - \sum_{IJ}\int\mathrm{d}^2\hat{z}_1\mathrm{d}^2\hat{z}_2\,\big[\varphi, \mathcal{F}_I(\hat{z}_1) \big] {W}_{IJ}(\hat{z}_1,\hat{z}_2) \big[\mathcal{F}_J(\hat{z}_2),\alpha\big] ~.
\end{align}
One immediate thing to notice is that\footnote{Note that after gauge fixing $\sqrt{q}=1$, $T_{ab}$ vanishes, and hence $\No_{ab} = -2\nabla_a\nabla_b\No{}^{\text{TF}}$.}
\begin{align}
    \big[ \mathcal{F}_{3ab}(\hx) \, , \, \alpha \big]_* = 0 \qquad \Rightarrow \qquad \big[\, \SNo_{ab}(\hx) \, , \, \alpha \big]_* = \big[ -2\nabla_a\nabla_b\No{}^{\text{TF}} \,,\, \alpha \big]_*  ~.
\end{align}
The only independent modes after imposing the constraints are $\{\No, C,N_{ab}\}$. One can notice that all of these modes have vanishing brackets with $\mathcal{F}_3$ and $\mathcal{F}_4$. Hence, only the $\mathcal{F}_1$, $\mathcal{F}_2$ block of the inverse matrix contributes to the correction of brackets. This particular block (upper left $2\times 2$ in \eqref{inverse MW}) is identically zero. Hence the brackets of these modes remain uncorrected:
\begin{align} \label{eq:phys_brackets}
      \big[N_{ab}(u,\hat{x})\,,\,N_{cd}(u',\hat{y})\big]_* &= \frac{1}{2} \,\p_u \delta(u-u')\, \sy_{ab,cd}\ \delta^2(\hx-\hy)\\
    \big[ C(\hx) \, , \, \No(\hy) \big]_* &= G(\hx,\hy)
\end{align}

\noindent These brackets will be useful in computing the Poisson algebra of the Weyl invariants in the next section. The structure of the gauged fixed phase space is not meant to imply that the reduced phase space is $\Gamma_{\text{HLMS}}$. We show in the next section that the supertranslations, which are a genuine symmetry of $\Gamma_{\text{HLMS}}$ are not well defined on the reduced phase space of the gauged Weyl model.

\subsubsection{Supertranslations in the presence of the gauged Weyl}

In this section, we show that the action of supetranslations is not well-defined on the reduced phase space.

Consider the action of supertranslations on the News tensor:
\begin{equation} \label{eq:sup_weyl_1}
    N_{ab}(u,\hx) \rightarrow N_{ab}(u+f(\hx),\hx)
\end{equation}
Since Weyl is a gauge redundancy, we may choose to act the same supertranslation on another representative of the orbit of $N_{ab}(u,\hx)$ under the Weyl scaling, say $N_{ab}(e^w u,\hx)$ :
\begin{equation}\label{eq:sup_weyl_2}
     N_{ab}(e^w u,\hx) \rightarrow N_{ab}(e^w u+f(\hx),\hx)
\end{equation}

\noindent As the right hand sides of \eqref{eq:sup_weyl_1}
and \eqref{eq:sup_weyl_2} are not related by a Weyl transformation, we conclude that the action of supertranslations take different points on the same gauge orbit to points that do not lie on the same gauge orbit, and are thus ill-defined on the reduced phase space. This is because the supertranslations and Weyl generating vector fields do not commute, see Weyl algebra from Section \ref{sec:Weyl} :
\begin{align}
\big[\delta_{(f,0,0)}\,,\, \delta_{(0,0,w)}\big] = \delta_{(wf,0,0)} ~.
\end{align}
Thus, if we gauge Weyl, supertranslations are no longer symmetries of the dynamical system. Since supertranslations are true symmetries of the quantum S-matrix, the gauged Weyl setup is unphysical. The absence of supertranslations as a symmetry manifests itself in the reduced phase space approach to gauging the Weyl action.

\subsubsection{Reduced Phase Space}
The Weyl action on $N_{ab}$ is induced by a scale transformation of the null coordinate. The algebra of dilatation invariant functions on $\mathbb{R}$  is isomorphic to the algebra of functions on a non-Hausdorff space consisting of three points. This non-trivial topology makes it clear why supertranslations no longer remain global symmetries in the reduced phase space.  

Let us now isolate the `physical' degrees of freedom without gauge fixing. To begin with, we have the following coordinates for the kinematical phase space:
\begin{align}
    \big\{\, \sqrt{q}\,,\,C(\hx)\,,\, \No(\hx) \,\big\} \cup \big\{ N_{ab}(u,\hx) \big\} ~.
\end{align}

The reduced phase space is the one constructed by endowing the space of Weyl-orbits a symplectic form induced from the kinematical phase space \cite{Gitman,sundermeyer}. Since each orbit may be parameterized by the values of the Weyl invariant quantities, we may treat independent Weyl invariant quantities as the coordinates on the reduced phase space. For instance, in the soft sector, kinematically, $\{C\,,\, \No\,,\, \sqrt{q}\}$ are the coordinates in the phase space. We can choose the following coordinates for the soft sector, in which case, the gauge mode and the physical modes are identifiable:
\begin{align}
     \big\{\, \sqrt{q}\,,\,\underbrace{\sqrt{q}^{-1/2}C\,,\, \sqrt{q}^{-1/2}\No}_{\text{Dynamical modes}} \,\big\} ~.
\end{align}
In addition to the $\{C,\No\}$ modes, we have the following integrals\footnote{Note that even though $u$ integrals, $\int_0^{\pm}\mathrm{d}u$, can be made Weyl invariant, they are not a part of the phase space since one needs distributional smearing functions to define them.} over $\scri$ that are naturally Weyl-invariant:
\begin{align}
    I^{\{a_n\}}_{\hdots} \equiv \sqrt{q}^{(\sum na_n-1)/2}\int_{-\infty}^\infty \mathrm{d}u\prod_{n=0}^\infty\big( \p_u^nN_{\cdot \cdot}(u) \big)^{a_n} \,, \qquad a_n \in \mathbb{Z}_{\geq 0} ~.
\end{align}
The $\hdots$ in $I^{\{a_n\}}_{\hdots}$ denotes the arbitrary sphere indices\footnote{One can verify the Weyl invariance of $I$ by using $\quad \p_u^nN_{ab}(u) \xrightarrow{w} e^{nw}\p_{u'}^nN_{ab}(u')\big|_{u'=e^wu}$.}. Note that we require only a finite number of $a_n$s to be non-zero for $I$ to be well defined. In holomorphic coordinates, these integrals take the following form:
\begin{align}
    I^{\{a_n\},\{b_n\}} \equiv \sqrt{q}^{(\sum n(a_n+b_n)-1)/2}\int_{-\infty}^\infty \mathrm{d}u\prod_{n=0}^\infty\big( \p_u^nN_{zz}(u) \big)^{a_n}\big( \p_u^nN_{\bar{z}\bar{z}}(u) \big)^{b_n} \,, \qquad a_n, b_n \in \mathbb{Z}_{\geq 0} ~.
\end{align}

Not all of the $I^{\{a_n \}}$ are independent - while some of them trivially vanish, others have relations amongst them arising out of integration by parts. However, such relations keep the rank of the tensor, $\sum na_n$ and $\sum a_n$ fixed. Examples include:

\begin{equation}
 q \int \mathrm{d}u \ N_{ab}N^{ab} \p_u N_{cd} +  2q\int \mathrm{d}u \ N_{ab} \p_u N^{ab}  N_{cd} = 0 ~.
\end{equation}

We can use the kinematical brackets to evaluate the physical Poisson brackets amongst Weyl invariant quantities. One can see clearly that the $I^{a_n}$ form a closed algebra: this is because, by virtue of \eqref{eq:phys_brackets}, the Poisson bracket of any two of these invariants is a sum over terms with one higher power of $\sqrt{q}$, one extra $u$ derivative. This ensures that the power of the prefactor $\sqrt{q}$ is always appropriate to make the Weyl weight of the overall quantity zero. Now we give a few examples of what the brackets look like. First, we note that the following set of observables forms an abelian subalgebra:
\begin{align}
    \left[ \sqrt{q}^{(4n-1)/2}\int\mathrm{d}u\left(N_{ab}N^{ab}\right)^n \,,\, \sqrt{q}^{(4m-1)/2}\int\mathrm{d}u'\left(N_{cd}N^{cd}\right)^m\right] = 0 ~.
\end{align}
For illustration, we write another Poisson bracket:
\begin{align}
    &\left[\sqrt{q}^{3/2}\int \mathrm{d}u\,N_{ab}\left(N_{mn}N^{mn}\right)(\hx)\,,\, \sqrt{q}^{2m-1/2}\int \mathrm{d}u'\left(N_{cd}N^{cd}\right)^m(\hy)\right] \nonumber\\&\hspace{4cm}= 2(1-m)\sqrt{q}^{2m}\int \mathrm{d}u\,\p_uN_{ab}\left(N_{cd}N^{cd}\right)^m\,\, \delta^2(\hx-\hy) ~.
\end{align}

The algebra of Poisson brackets amongst $I^{\{a_n\}}$ is compatible with the Poisson algebra derived by fixing the gauge, \eqref{eq:phys_brackets}.  

We can act arbitrary Diff-$S^2$ derivatives on any of these Weyl invariant quantities, and with appropriate factors of $\sqrt{q}$, we obtain more Weyl invariant tensors. The reduced phase space is thus parameterized by the soft sector and the integrals $I^{\{a_n\}}$. These  constitute an over-complete set of invariants, and characterizing the reduced phase space by picking out a basis from this set of invariants will be a challenging problem. 

\section{Conclusions}
The recent studies in asymptotic symmetries beyond the ones generated by BMS vector fields typically relies on the relaxed boundary conditions in which the celestial sphere metric is allowed to fluctuate. However, our experience with asymptotic quantization of radiative phase spaces in which the celestial metric is dynamical is rather limited \cite{am}. This is because passing to quantum theory requires that we find the conjugate pairs in the soft and the hard sector. The broad goal in this paper has been to provide examples of such radiative phase spaces, which are in principle amenable to asymptotic quantization. We have analysed two scenarios in which the sphere metric is dynamical: gBMS under the restricted setting of linearized gravity and Weyl BMS where we treat the Weyl scalings as pure gauge.

We obtained the physical radiative phase space for gBMS in the context of linearized gravity. Even though the generic Dirac bracket involves an abstract Green's function, we have identified a certain combination of Geroch tensor and celestial metric that is conjugate to the sub-leading soft News tensor. The final phase space does not factorize neatly into the hard and soft sectors, as the brackets amongst these soft modes and hard modes are non-vanishing. The results of this analysis parallel those of \cite{am}, indicating the robustness of those results. 

In the linearized gravity approach to the gBMS phase space, we showed that keeping Geroch tensor $T_{ab}$ as a constant mode of the News tensor leads to simplifications in our analysis. A similar approach may help in the non-linear analysis as well.  

We showed that the usual method of obtaining Poisson brackets from the symplectic form via Dirac's method of eliminating second class constraints runs into some issues when applied to the symplectic form for gBMS in \cite{mp}. The core obstruction to the program is the fact that the determinant of the Dirac matrix is an operator that has functional dependence on soft modes. We believe that it will require new ideas to solve the constraints in this case as an inversion of an operator-valued Dirac matrix appears to be a rather intractable problem.  Given the importance of obtaining the radiative phase space to perform (asymptotic) quantization and define a complete set of soft vacua in quantum gravity, the 
problem merits a serious investigation. 

We then focused on a toy model involving the Weyl BMS group but excluding the superrotation subgroup. Restricting our attention to the special case with vanishing Weyl flux, we obtained a symplectic form that is degenerate over the orbit of the pure Weyl transformations of the Weyl BMS group. We checked that the \emph{Christodoulou-Klainerman (CK)} condition is invariant under the action of the Weyl rescaling and obtained the reduced phase space for the gauged Weyl model. It is an intriguing observation that the reduced phase space of the HLMS phase space augmented with Weyl scalings as gauge, is not HLMS anymore, as it lacks supertranslations. 

The gauge invariant observables are precisely those that are fixed under Weyl rescalings. In the soft sector, this transformation acts via an overall scaling and thus we have specified a generating set of gauge invariant quantities by supplementing the soft modes by appropriate factors of sphere area element $\sqrt{q}$. However, in the hard sector, there is no notion of local hard News and the supertranslations are not well defined. As a result, Weyl scalings can not be a gauge redundancy in any physical setting. Also, it leads to the fact that in the hard sector, only the Weyl invariant densities constructed out of the $\scri$ integrals of hard News survive in the reduced phase space. We have identified a set of such quantities that form a closed algebra, and presumably, form an overcomplete set of Weyl invariants. 

It would be natural to explore how our work in both settings connects with the flat space holography program (see \cite{Pasterski2016,ccft} for a review). 

The reduced phase space defined in the gauged Weyl setting may have an interesting connection with the space of all asymptotically locally flat geometries which were analyzed in \cite{nishant}. The phase space description we provide in this work may prove useful to better understand these locally flat geometries.
  
Our reduced phase space analysis for linearized gBMS could prove useful in a variety of settings. For instance, the double soft graviton theorems are sensitive to the details of charge algebras \cite{distler fh, anupam kundu}. It would be useful to revisit the double soft theorems in light of conservation laws emerging out of the gBMS phase space. 

\acknowledgments

We are grateful to Alok Laddha for suggesting the problem and thank him for many fruitful discussions and constant encouragement. We would like to thank Sujay Ashok for the discussions and for providing feedback on the draft. We thank Miguel Campiglia, Laurent Freidel, and Daniele Pranzetti for the helpful correspondence.  We also appreciate the long and helpful discussions with PB Aneesh, Nishant Gupta, and Dileep Jatkar over the past year.

\appendix

\section{Dirac Analysis} 

\noindent Given a dynamical system and the corresponding Poisson brackets, if we wish to impose some constraints $F_k = 0$ on the dynamical variables, then we follow Dirac's analysis. (We refer to \cite{d} for a comprehensive review.) The brackets of the dynamical quantities on the constrained surface are called \emph{Dirac brackets}. Given the Poisson brackets amongst the dynamical variables, we can find brackets amongst the constraints as well. If a constraint commutes with all the other constraints (valued on the constrained surface), then it is called a first-class constraint. Constraints that are not first class are termed second class. 

Given a set of constraints, consider the maximal set of second-class constraints. Let us denote them by $\mathcal{F}_k$. We denote the kinematical brackets as $\big[\cdot\, ,\, \cdot \big]$, and the Dirac brackets as $\big[\cdot\, ,\,\cdot \big]_*$. The prescription due to Dirac for computing these new brackets is  :
\begin{align}
    \big[\varphi,\alpha\big]_* = \big[\varphi,\alpha\big] - \sum_{i,k}\big[\varphi, \mathcal{F}_i \big] \mathcal{W}^{ik} \big[\mathcal{F}_k,\alpha\big]  \ .
\end{align}
Here $\varphi$ and $\alpha$ are arbitrary functions on phase space and $\mathcal{W}^{ik}$ is the inverse of the commutator of constraints, defined as follows :
\begin{align}
\sum_{i}\big[\mathcal{F}_l\,,\, \mathcal{F}_i\big] \mathcal{W}^{ik} = \delta^k_l &= \sum_i\mathcal{W}^{ki}\big[\mathcal{F}_i,\mathcal{F}_l \big] ~.\\
    \text{Note that} \hspace{2cm} \big[ \cdot , \mathcal{F}_i \big]_* &= 0  \ .
\end{align}

\noindent When the phase space is finite-dimensional, one can check that the Jacobi identity is satisfied. Hence the final brackets are antisymmetric and satisfy the Jacobi identity by construction. 

The above construction can be generalized to the continuous case. In that case, the sum over constraints gets additional integrals appropriately as follows: 
\begin{align}
    \sum_K \int \mathrm{d}^2\hat{z}\,\big[ \mathcal{F}_I(\hx) \,,\, \mathcal{F}_K(\hat{z}) \big]\,W^{KL}(\hat{z},\hy) = \delta^L_I\,\delta^2(\hx-\hy) ~.
\end{align}
Recall that the antisymmetry of the Dirac matrix manifests itself as follows:
\begin{align}
   M_{IK}(\hx,\hat{z}) \equiv  \big[ \mathcal{F}_I(\hx) \,,\, \mathcal{F}_K(\hat{z}) \big] = - \big[ \mathcal{F}_K(\hat{z}) \,,\, \mathcal{F}_I(\hat{x}) \big] = - M_{KI}(\hat{z},\hx) ~.
\end{align}
The same holds for the inverse as well.

\section{Dirac Matrix in the gBMS Case}
\label{appendix B}

Let us reiterate the symplectic form for the gBMS case \eqref{eq:gBMS_symp}:\footnote{Note that the Geroch tensor is explicitly present, unlike the case of Section \ref{linearized section}. This is due to $\Pi^{ab}$ being intricate in the non-linear case and there being no canonical way to incorporate $\delta\Pi \wedge \delta T$ into the hard sector.}
\begin{align}
    \Omega &=  \int_{\mathcal{I}} \sqrt{q} \ \delta N_{ab} \wedge \delta \overset{o}{\sigma}{}^{ab} + \int_{S^2}  \sqrt{q} \ \left( \delta \overset{o}{N} \wedge \ \delta \sC + \delta \Pi^{ab}\wedge \delta T_{ab} +  \delta p^{ab} \wedge \delta q_{ab} \right) ~,\\
    & \quad \text{with} \qquad \sC = \left(-2\nabla_a\nabla_b + q_{ab}\Delta +T_{ab}\right)\left(-2\nabla^a\nabla^b + T^{ab}\right) C \label{relation sc C} ~.
\end{align}

We have traded off $C$ in favor of $\sC$. Given $\sC$, we can invert the differential operator and obtain $C$ in terms of the appropriate Green function. The information regarding the boundary modes of shear is completely encoded in $\sC$, rather than $C$. After finding the brackets of 
$\sC$, we can use \eqref{relation sc C} to obtain the brackets of $C$ as well. 

The variables $p^{ab}$ and $\Pi^{ab}$ are both tracefree. This is because their conjugates also have two degrees of freedom each, as they satisfy $\text{det } q_{ab} = 1$ and $q^{ab}T_{ab}=0$. One can see that by going to the local complex coordinates. In abstract index notation, since $q_{ab}$ is dynamical, the tracelessness condition itself is intricate, as can be seen in the following illustration:
\begin{align}
    \delta p^{ab} \wedge \delta q_{ab} &= \delta p^{ab}{}^{\text{TF}}\wedge \delta q_{ab} + \frac{1}{2}\delta (p\, q^{ab}) \wedge \delta q_{ab} \\
    &= \delta p^{ab}{}^{\text{TF}}\wedge \delta q_{ab} + \frac{1}{2}\left( \delta p \wedge q^{ab}\delta q_{ab} + p\,\delta q^{ab}\wedge \delta q_{ab}\right) \\
    &= \delta p^{ab}{}^{\text{TF}}\wedge \delta q_{ab} ~.\\
  \text{Where we used}\quad\qquad   q^{ab}\delta q_{ab} &\sim \delta \sqrt{q} = 0 \ \ , \qquad \delta q^{ab} \wedge \delta q_{ab} \sim \delta (\delta \sqrt{q}) = 0~. \nonumber
\end{align}

To obtain the kinematical brackets, we calculate the HVFs.\footnote{Note that in the following, we treat $\So_{ab}$ as the fundamental field and $\So{}^{ab}$ as a functional of $\So_{ab}$ and $q_{ab}$.}
\begin{align}
     X_{N_{ab}} &= \frac{1}{2}\, \sy_{ab,mn}\,\frac{\delta}{\delta  \overset{o}{\sigma}_{mn}} -  \, N^n{}_a \frac{\delta}{\delta p^{nb}} \ ,\\
     X_{\SNo_{ab}} &= \sy_{ab,mn}\,\int \mathrm{d}u\,\frac{\delta}{\delta  \So_{mn} } \ ,\\
     X_{\SNi{}^{ab}} &= \frac{1}{2}\int\mathrm{d}u\,u\,\frac{\delta}{\delta  \So_{mn} } + q^{am}\SNi{ }^{bn}\frac{\delta}{\delta p^{mn}} \ ,\\
     X_{T_{ab}} &= -\frac{\delta}{\delta \Pi^{ab}}\ , \hspace{3cm} \\X_{\Pi^{ab}} &= \frac{\delta}{\delta T_{ab}} \ ,\\
X_{\sC} &= -\frac{\delta}{\delta   \No} \ ,\hspace{3.2cm}\\ X_{\No} &= \frac{\delta}{\delta \sC} \label{eq:C_No} \ ,\\
X_{q_{ab}} &= -\frac{\delta}{\delta p^{ab}}  \ ,\hspace{3cm} \\X_{p^{ab}} &= \frac{\delta}{\delta q_{ab}} + 2\int_{\mathcal{I}}q^{(ma}\So{ }^{bn)} X_{N_{mn}}\ . 
\end{align}

\noindent The relevant kinematic brackets among the phase space variables follow from the HVFs, and here we state the results :
\begin{align}
    \big[  \So_{ab}(u,\hat{x})\,,\, N_{cd}(u',\hat{y}) \big] &= \frac{1}{2}\delta(u-u')\,\sy_{ab,cd}\,\delta^2(\hx-\hy) \ ,\\
    \big[ \So_{ab}(u,\hx)\,,\, \SNo_{cd}(\hy)\big] &= \sy_{ab,cd} \ \delta^2(\hx-\hy) \label{eq:F3F4}\ , \\
    \big[ \So_{ab}(u,\hx)\,,\, \SNi{ }^{cd}(\hy)\big] &= \frac{1}{2} u\,  \sy^{cd}_{ab} \ \delta^2(\hx-\hy) \ ,\\
    \big[ N_{ab}(u,\hx)\,,\,p^{cd}(\hy) \big] &= N_a^{(c}\delta^{d)}_b\,\delta^2(\hx-\hy)\ ,\\
    \big[T_{cd}(\hx)\,,\, \Pi^{ab}(\hy)\big] &=  \sy^{ab}_{cd} \ \delta^2(\hx-\hy) \ , \\
    \big[q_{cd}(\hx)\,,\, p^{ab}(\hy)\big] &=  \delta^{ab}_{cd} \ \delta^2(\hx-\hy) \ ,\\
    \big[p^{cd}(\hx)\,,\, \SNi{ }^{ab}(\hy)\big] &=  \left(\,\SNi{ }^{b (d} q^{a c) }-\frac{1}{2}q^{cd}\SNi{ }^{ab}\right)\delta^2(\hx-\hy)\ .
\end{align}

The kinematical structure is thus well understood. We now have to impose the constraints \eqref{eq:p} - \eqref{f5} on this kinematical phase space. We can compute the kinematical brackets of the constraints yielding the Dirac matrix. At generic points in the phase space, these constraints are rather unwieldy, so as a first step we consider the region $\Gamma_{ts}$ defined as the subspace of the phase space where the modes $\No$ and $\SNi_{ab}$ vanish. In this special case, we note that the brackets of the bi-linear terms in $C$ and $\No$ vanish, and hence these terms have no bearing on what follows.\footnote{In addition to all the brackets that vanish, we have omitted any bracket involving the modes $C$ and $\No$. This is because by virtue of \eqref{eq:C_No} the only non-zero bracket the soft mode $\No$ has is the $[C,\No]$ bracket. But since in the constraints, $C$ and $\No$ always come in pairs, all the brackets involving these vanish when $\No$ is set to zero. Thus we may ignore the terms from the constraints that are bi-linear in $C$, $\No$.} We then show that the determinant of the resulting matrix is zero. This indicates that the determinant of the Dirac matrix is functionally dependent on the News tensor, unlike the case in \cite{am}.  

Although the precise elements of the Dirac matrix can be computed using these kinematical brackets, these expressions are cumbersome and hence are omitted from this document. For our purposes here, we merely need to keep track of whether the element is non-zero or not, on the points where $\SNi_{ab}$ and $\No$ are set to zero. Crucial to our results will be the vanishing of three kinematical brackets : $[p^{ab},p^{cd}] = [p^{ab},\Pi^{cd}] = [\SNi{ }^{ab},\SNi{ }^{cd}] =0$.  

We begin with $[\mathcal{F}_1^{ab},\mathcal{F}_1^{cd}]$. Notice that every term in $\mathcal{F}_1$ (modulo  expressions involving the $C$ mode) has an explicit $\SNi{ }^{ab}$ in it. Since the only non-vanishing bracket involving $\SNi{ }^{ab}$, of the form $\big[p^{cd}, \SNi{ }^{ab}\big]$, happens to be proportional to $\SNi{ }^{cd}$, we deduce that $[\mathcal{F}_1^{ab},\mathcal{F}_1^{cd}]$ vanishes when the sub-leading News tensor is set to zero.

The story is similar with $[\mathcal{F}_1^{ab},\mathcal{F}_2^{cd}]$ : since $[p^{ab},\Pi^{cd}]$ and $[\SNi{ }^{ab},\SNi{ }^{cd}]$ vanish, the leftover bracket is precisely $\big[p^{cd}, \SNi{ }^{ab}\big]$, which is proportional to $\SNi{ }^{ab}$. Of the remaining brackets, note that $[(\mathcal{F}_5)_a,(\mathcal{F}_5)_b]$ and $[\mathcal{F}_2^{ab},\mathcal{F}_2^{cd}]$ vanish (modulo terms with the $C$ mode). The brackets $[\mathcal{F}_1^{ab},(\mathcal{F}_5)_c]$ and $[\mathcal{F}_2^{ab},(\mathcal{F}_5)_c]$ are independent of $\SNi{ }^{cd}$ and $\No$. Their precise form will not be required for our arguments. The last brackets are those involving $\mathcal{F}_3$  and $\mathcal{F}_4$ : in this case, the condition $\No =0 $ implies that these constraints decouple from the rest and the Dirac matrix block diagonalizes. The $\mathcal{F}_3$,$\mathcal{F}_4$ block is invertible and is functionally independent of the shear tensor, see \eqref{eq:F3F4}. The following is a block of the Dirac matrix, formed by constraints $\mathcal{F}_1$, $\mathcal{F}_2$, and $\mathcal{F}_5$. 

\begin{align}
\left(\begin{matrix}
\big[ \mathcal{F}_1 \,,\, \mathcal{F}_1 \big] & \big[ \mathcal{F}_1 \,,\, \mathcal{F}_2 \big] & \big[ \mathcal{F}_1 \,,\, \mathcal{F}_5 \big] \\
\big[ \mathcal{F}_2 \,,\, \mathcal{F}_1 \big] & \big[ \mathcal{F}_2 \,,\, \mathcal{F}_2 \big] & \big[ \mathcal{F}_2 \,,\, \mathcal{F}_5 \big] \\
\big[ \mathcal{F}_2 \,,\, \mathcal{F}_1 \big] & \big[ \mathcal{F}_5 \,,\, \mathcal{F}_2 \big] & \big[ \mathcal{F}_5 \,,\, \mathcal{F}_5 \big]\\
\end{matrix}\right) = \left(\begin{matrix}
\left(\hdots \SNi \hdots\right) & \left(\hdots \SNi \hdots\right) & \left( \hdots \right) \\
\left(\hdots \SNi \hdots\right) & 0 & \left( \hdots\right) \\
\left( \hdots \right) &  \left( \hdots \right) & 0\\
\end{matrix}\right) \label{full gbms problem}
\end{align}

\noindent Each entry in this $3\times 3$ matrix is a $2\times 2$ block. In the matrix shown in \eqref{full gbms problem}, $\hdots \SNi \hdots $ denotes terms that are linear in subleading soft News and vanish when it is set to zero. Note that $\No$ has already been set to zero. If $\SNi{ }^{ab}$ is also zero, the top four rows of the Dirac matrix form a set of 4 vectors, each of which is $2$ dimensional. Thus, the  determinant of this matrix must vanish. On the other hand, one can explicitly check that determinants of matrices with non-zero entries in the locations represented by $ \hdots \SNi \hdots $ do not trivially vanish. Thus we conclude that the determinant of the Dirac matrix is functionally dependent on the Bondi News tensor, as claimed in Section \ref{obstacles section}.

The aforementioned determinant also happens to be a differential operator (acting on the Dirac delta function of $S^2$). Since the inverse matrix involves the inverse of the determinant, we would expect it to involve Green's function for this operator. The problem of finding this Green's function, however, is intractable, as the operator is dependent on the soft modes means that its coefficients are arbitrary functions on the celestial sphere. We thus see that, one can not solve the Dirac constraints globally on the phase space and that they can only be solved locally in phase space (for fixed soft modes.) However, this analysis is outside the scope of the present paper. Though, the extended covariant phase space of linearized theory that includes (in addition to the canonical phase space of linearized gravity) the radiative phase space of leading and sub-leading soft modes turns out to be tractable, as described in Section \ref{linearized section}.

The entries of the inverse of the Dirac matrix, if it exists, are precisely the functionals of physical brackets among the soft sector modes. For instance consider the brackets $\big[q_{ab}(\hx),q_{cd}(\hy)\big]_{*}$. (Kinematically, $q_{ab}$ and $T_{ab}$ only have non-zero brackets with $p^{ab}$ and $\Pi^{ab}$ respectively.)

\begin{align}
    \big[q_{ab}(\hx),q_{cd}(\hy)\big]_{*} &= - \int\mathrm{d}^2\hat{z}_1\mathrm{d}^2\hat{z}_2\, \big[q_{ab}(\hx),\mathcal{F}^{ij}_1(\hat{z}_1)\big]\, W^{11}{}_{ij,kl}(\hat{z}_1,\hat{z}_2) \,\big[\mathcal{F}^{kl}_1(\hat{z}_2),q_{cd}(\hy)\big] \\
    &= - \int\mathrm{d}^2\hat{z}_1\mathrm{d}^2\hat{z}_2\, \big[q_{ab}(\hx),p^{ij}(\hat{z}_1)\big]\, \mathcal{W}^{11}{}_{ij,kl}(\hat{z}_1,\hat{z}_2) \,\big[p^{kl}(\hat{z}_2),q_{cd}(\hy)\big] \\
    &= \mathcal{W}^{11}{}_{ab,cd}(\hat{x},\hat{y}) \ .
\end{align}
Similarly, for $T_{ab}$, we have:
\begin{align}
    \big[ T_{ab}(\hx),T_{cd}(\hy) \big]_* = \mathcal{W}^{22}_{ab,cd}(\hx,\hy) \ .
\end{align}

Thus, finding the physical brackets is about as hard a problem as solving the set of coupled PDEs coming from the constraint analysis.

\bibliographystyle{JHEP}

\end{document}